\documentclass[onecolumn]{aa}
\usepackage{graphicx}
\usepackage{color}
\usepackage{txfonts}

\begin{document}

\title{Rossby waves and polar spots in rapidly rotating stars: Implications for stellar wind evolution}

\author{Zaqarashvili, T.V.\inst{1,5}, Oliver, R.\inst{2}, Ballester, J.L.\inst{2}, Carbonell, M.\inst{3},  Khodachenko, M.L. \inst{1}, Lammer, H.\inst{1},\\
Leitzinger, M.\inst{4},  and Odert, P. \inst{4}}

 \institute{ Space Research Institute, Austrian Academy of Sciences, Schmiedlstrasse 6, 8042 Graz, Austria\\
             \email{[teimuraz.zaqarashvili;maxim.khodachenko;helmut.lammer]@oeaw.ac.at}
                               \and
             Departament de
F\'{\i}sica, Universitat de les Illes Balears, E-07122 Palma de
Mallorca, Spain\\
             \email{[ramon.oliver;joseluis.ballester]@uib.es}
                                 \and
                                 Departament de Matem\`{a}tiques i Inform\`{a}tica, Universitat de les Illes Balears, E-07122 Palma de
Mallorca, Spain\\
             \email{[marc.carbonell]@uib.es}
                                 \and
             Institut f\"ur Physik, IGAM, Karl-Franzens Universit\"at Graz, Universit\"atsplatz 5, A-8010 Graz, Austria\\
             \email{[martin.leitzinger;petra.odert]@uni-graz.at}
                                \and
            Abastumani Astrophysical Observatory at Ilia State University, University St. 2, Tbilisi, Georgia\\
}

\date{Received / Accepted }

\abstract{Rapidly rotating stars show short-period oscillations in
magnetic activity and polar appearance of starspots.}{The aim of
this paper is to study large-scale shallow water waves in the
tachoclines of rapidly rotating stars and their connection to the
periodicity and the formation of starspots at high
latitudes.}{Shallow-water magnetohydrodynamic equations were used to
study the dynamics of large-scale waves at the rapidly rotating
stellar tachoclines in the presence of toroidal magnetic field.
Dispersion relations and latitudinal distribution of wave modes were
derived.}{We found that low-frequency magnetic Rossby waves tend to
be located at poles, but high-frequency magnetic Poincar\'e waves
are concentrated near the equator in rapidly rotating stars. These
results have important implications for the evolution of the stellar
wind in young Sun-like stars.}{Unstable magnetic Rossby waves may
lead to the local enhancement of magnetic flux at high latitudes of
tachoclines in rapidly rotating stars. The enhanced magnetic flux
may rise upwards owing to the magnetic buoyancy in the form of tubes
and appear as starspots at polar regions. Magnetic Rossby waves may
also cause observed short-term periodicity in the stellar magnetic
activity. }


\titlerunning{Rossby waves in rapidly rotating stars}

\authorrunning{Zaqarashvili et al.}

\maketitle

\section{Introduction}

Rotation and convection lead to stellar dynamo activity, which
results in the strong concentration of magnetic fields on the
surface. These magnetic features are called {\it starspots}, and
they represent the sunspot analog for other stars. Interested
readers may find detailed information about starspots in two recent
comprehensive reviews: Berdyugina (\cite{Berdyugina2005}) and
Strassmeier (\cite{Strassmeier2009}). Hall (\cite{Hall1972})
explicitly postulated the starspot model as explaining the wave-like
features in the light curves of rotating RS CVn stars. Doppler
imaging technique showed spots on the RS CVn star HR 1099, and they
exhibit a similarity in shape and location to X-ray images of solar
coronal holes (Vogt and Penrod, \cite{Vogt1983}). Later, Strassmeier
(\cite{Strassmeier1990}) observed a cool polar spot on the rapidly
rotating G5 subgiant in the single-lined RS CVn binary HD 26337. The
spot was rather long-lived, persisting for at least the nine years
of observations.

Zeeman-Doppler imaging of the circularly polarized spectrum has led
to the first magnetic maps of the active K1 subgiant in the RS CVn
binary system HR 1099 = V711 Tau (K1IV+G5V) other than the Sun,
which also showed the polar location of starspots (Donati et al.
\cite{Donati1992}). It was generally accepted that starspots on
magnetically active, cool stars mainly appear near the poles.
Sch\"ussler and Solanki (\cite{Schussler1992}) suggested that the
preference for high latitudes is due to the rapid rotation to these
stars, which in turn leads to a dominance of the Coriolis force over
the buoyancy force in the dynamics of magnetic flux tubes.
Later, Sch\"ussler et al. (\cite{Schussler1996}) carried out
simulations on solar-type stars in an effort to predict the
emergence latitude of magnetic flux tubes, after evolution from a
toroidal magnetic field located at the base of the convection zone.
They carried out simulations for stars of different ages and
equatorial rotation velocities. The results predict high-latitude
spots since, in a fast rotator, Coriolis forces acting on a rising
magnetic flux tube cause movement parallel to the rotation axis.

On the other hand, Barnes et al. (\cite{Barnes1998}) studied two
rapidly rotating G dwarfs in the $\alpha$ Persei cluster, from
spectra taken at the William Herschel Telescope. The image
reconstructions demonstrate that both these stars have cool
high-latitude regions or polar crowns and low-latitude features,
contradicting the suggestion that only high-latitude spots should be
present. The magnetic topology of AB Dor (young K0 dwarf) is also
found to be very complex, with at least 12 different radial field
regions of opposite polarities located all around the star (Donati
et al. \cite{Donati1999, Donati2003}). Significant azimuthal field
fluxes are also detected in the form of one negative polarity region
close to the equator, a series of positive polarity patches at
intermediate latitudes and an almost complete ring of negative
polarity encircling the rotational pole at high latitudes. Starspots
on the young solar analog EK Dra (rotation period of 2.68 days) are
observed to be formed everywhere, from the low latitudes up to polar
regions (Strassmeier 2009). Therefore, the current view is that
starspots appear at all latitudes in rapidly rotating stars.

The high-latitude sunspots could change the structure of coronal
magnetic field as well as the formation of prominences, which may
lead to the predominance of solar flares and coronal mass ejections
(CMEs) near the polar regions. CMEs are believed to have tremendous
influence on the atmosphere of exoplanets, leading to their erosion
(Khodachenko et al. \cite{Khodachenko2007}). The polar CMEs may
propagate significantly out off the ecliptic plane, which may cause
more convenient conditions of planetary evolution; therefore, the
detailed study of starspot distribution in rapidly rotating young
solar analogs is very important for planetary evolution as well.

Observation of starspots at all latitudes complicates any
explanation of their appearance. There are at least two suggested
mechanisms of starspot formation in rapidly rotating stars. The
first mechanism, which we already mentioned above, is that the
magnetic tubes tend to move to the polar regions owing the Coriolis
force acting during the rising across the convection zone
(Sch\"ussler and Solanki \cite{Schussler1992}; Sch\"ussler et al.
\cite{Schussler1996}; DeLuca et al. \cite{DeLuca1997}; Granzer et
al. \cite{Granzer2000}; Holzwarth et al. \cite{Holzwarth2006}, Isik
et al. \cite{Isik2007}). The second mechanism is the transport of
photospheric magnetic flux towards the polar regions by meridional
advection  (Schrijver and Title \cite{Schrijver2001}, Mackay et al.
\cite{Mackay2004}). The first mechanism may explain the starspot
appearance at high latitudes, but it may fail to explain the
equatorial spots because the magnetic tubes should all tend to
higher latitudes in rapid rotators. On the other hand, the second
mechanism may explain the distribution of spots at all latitudes,
but it requires a strong meridional flow in order to carry the
magnetic flux towards the poles.

Another interesting observational indication is the cyclic behavior
of stellar activity with shorter periods than the solar cycle
($\sim$ 11 yr). Massi et al. (\cite{massi1998}, \cite{massi2005})
report short-term activity cycles on UX Arietis, which is a binary
system formed by a spotted K0 IV primary and a G5 V secondary with
an orbital period of 6.44 days. Short-period activity cycles have
been observed on $\tau$ Bootis (Donati et al. \cite{Donati2008}).
Ol\'ah et al. (\cite{Olah2009}) analyzed 20 active stars and came to
the conclusion that stellar activity cycles are generally multiple
and variable. Recently, Lanza et al. (\cite{Lanza2009}) have
observed a short-term cyclic oscillation of the total spotted area
on the photosphere of CoRoT-Exo-2a, which is supposedly caused by
Rossby waves. The rotation period of CoRoT-Exo-2a is 4.522 $\pm$
0.024 days, while the total spotted area shows a cyclic oscillation
with a period of 28.9 $\pm$ 4.3 days. Almost all observed
periodicities are shorter than the solar cycles and are several
times longer than the stellar rotation periods. It should be
mentioned, however, that solar activity also undergoes intermediate
periodicities in the range of several months (Rieger et al.
\cite{Rieger84}, Carbonell et al. \cite{Carbonell1990}, Oliver et
al. \cite{oliver1998}) and $\sim$ 2 years (Sakurai \cite{sakurai81},
Gigolashvili et al. \cite{gigolashvili95}, Vecchio et al.
\cite{vecchio10}). These periodicities are explained in terms of
magnetic Rossby wave instability in the solar tachocline that is the
joint effect of latitudinal differential rotation and toroidal
magnetic field (Zaqarashvili et al. \cite{Zaqarashvili2010a},
\cite{Zaqarashvili2010b}). The coincidence between the solar short
term oscillations and the cyclic behavior of spotted stars indicate
that magnetic Rossby waves can be important in rapidly rotating
stars.

Here we study the behavior of magnetic shallow water waves in
tachoclines of rapidly rotating stars and their role in the
latitudinal distribution of starspots. The tachocline is the thin
layer located between the external convection envelope and the
internal radiative core (in the case of the Sun see Spiegel and Zahn
\cite{Spiegel1992}). Rossby waves are well studied in the
geophysical context (Gill \cite{gill82}); however, the presence of
magnetic fields significantly modifies their dynamics (Zaqarashvili
et al. \cite{Zaqarashvili2007}, \cite{Zaqarashvili2009}, Heng and
Spitkovsky \cite{Heng2009}, Mathis and de Brye \cite{mathis2011}).
We have used the shallow water magnetohydrodynamic (MHD) equations
(Gilman \cite{Gilman2000}) to study the latitudinal structure of
magnetic shallow water waves in tachoclines of rapidly rotating
stars. We first write the governing equations of shallow water MHD
in the rotating frame. Then, we derive the dispersion relations and
latitudinal structure of magnetic shallow water waves in different
configurations of toroidal magnetic field, which is followed by the
discussion of the connection between the shallow water waves and the
starspots at stellar photospheres.

Finally we discuss the implications of our findings in the context
of the evolution of young stellar winds of Sun-like stars. There is
observational evidence that zero-age-main-sequence solar-type stars
rotate more than ten times faster than the present Sun (e.g.,
G\"udel et al. \cite{Gudel1997}, Ribas et al. \cite{Ribas2005},
Lammer et al. \cite{Lammer2011}). As a consequence, young solar-type
GV stars, including the early Sun, should have vigorous magnetic
dynamos and correspondingly strong high-energy emissions. From the
study of solar-type stars with different ages, Skumanich
(\cite{Skumanich1972}), Simon et al. (\cite{Simon1985}) and others
have shown that these young stars lose angular momentum with time
via magnetized stellar winds, thus leading to a secular increase in
their rotation period (Durney \cite{Durney1972}). This rotation
slowdown is well fitted by a power law roughly proportional to
$t^{1/2}$, where $t$ is the  stellar age in units of Gyr (e.g.,
Skumanich \cite{Skumanich1972}; Soderblom \cite{Soderblom1982};
Ayres \cite{Ayres1997}). As a response to slower rotation the solar
dynamo strength diminishes with time, causing the Sun's high-energy
emissions also to undergo a significant decrease (Ribas et al.
\cite{Ribas2005}). The fundamental question of whether the young
fast-rotating Sun and similar stars have experienced some periods of
stronger magnetic activity has a major impact on the evolution and
plasma interaction of planetary atmospheres in the Solar System and
beyond.

\section{Governing equations}

We use a spherical coordinate system $(r, \theta, \phi)$  rotating with the stellar equator, where $r$ is the
radial coordinate, $\theta$ is the co-latitude, and $\phi$ the
longitude. We consider an unperturbed toroidal magnetic field $B_{\phi}=B({\theta})
\sin{\theta}$. Then,
the linearized form of shallow-water MHD equations can be rewritten in the rotating frame as (Zaqarashvili et al. \cite{Zaqarashvili2007}, \cite{Zaqarashvili2010a})
\begin{eqnarray}\label{eq:21}
{{\partial {\hat u}_{\theta}}\over {\partial t}} - 2\Omega_0 \cos
\theta {\hat u}_{\phi} +{{g}\over R_0}\sin \theta{{\partial h}\over
{\partial \theta}}- {{B({\theta})}\over {{4\pi\rho R_0}}}{{\partial {\hat
b}_{\theta}}\over {\partial \phi}}+ 2{{B({\theta})}\over
{{4\pi\rho R_0 }}}{\cos \theta}{\hat b}_{\phi}=0, \end{eqnarray}
\begin{equation}\label{eq:22}
{{\partial {\hat u}_{\phi}}\over {\partial t}} + 2\Omega_0 \cos
\theta {\hat u}_{\theta} +{{g}\over {R_0}}{{\partial h}\over
{\partial \phi}}- {{{\hat b}_{\theta}\sin \theta}\over {{4\pi\rho R_0}}}{{\partial
B({\theta})}\over {\partial \theta}}-{{B({\theta})}\over {{4\pi\rho R_0}}}{{\partial {\hat
b}_{\phi}}\over {\partial \phi}}-{{2B({\theta})}\over {{4\pi\rho R_0
}}}{\cos \theta}{\hat b}_{\theta}=0,
\end{equation}
\begin{equation}\label{eq:23}
\sin^2 \theta{{\partial h}\over {\partial t}} + {{H_0}\over {R_0
}}\sin \theta{{\partial {\hat u}_{\theta}}\over {\partial \theta}} +
{{H_0}\over {R_0}}{{\partial {\hat u}_{\phi}}\over {\partial
\phi}}=0,
\end{equation}
\begin{equation}\label{eq:24}
{{\partial {\hat b}_{\theta}}\over {\partial t}}- {{B({\theta})}\over
{{R_0}}}{{\partial {\hat u}_{\theta}}\over {\partial \phi}}=0,
\end{equation}
\begin{equation}
{{\partial {\hat b}_{\phi}}\over {\partial t}}+ {{1}\over
{{R_0}}}\sin \theta{{\partial} \over {\partial
\theta}}[{\hat u}_{\theta} B({\theta})]=0,
\end{equation}
where ${\hat u}_{\theta}= \sin{\theta}u_{\theta}$, ${\hat u}_{\phi}=
\sin{\theta}u_{\phi}$, ${\hat b}_{\theta}= \sin{\theta}b_{\theta}$, and ${\hat b}_{\phi}= \sin{\theta}b_{\phi}$ are the
velocity and magnetic field perturbations, $H_0$ is the
tachocline thickness, $h$ the
perturbation of the thickness, $\rho$ the density, $R_0$ the distance
from the stellar center to the tachocline,  $g$ the reduced gravity (due to subadiabatic temperature gradient; Gill \cite{gill82}) and $\Omega_0$ the equatorial angular velocity. For a zero magnetic field, this system transforms into the HD shallow-water equations.

We now perform a Fourier analysis of the form $\exp(-i\omega t
+ is\phi)$ and define
\begin{eqnarray}\label{eq:variables}
{{\omega}\over {2\Omega_0}}=\lambda,\,{{4\Omega^2_0R^2_0}\over {g
H_0}}=\epsilon,\, {{B^2({\mu})}\over {4\pi \rho}}=v^2_A(\mu),\,{{v_A(\mu)}\over {2\Omega_0R_0}}=\alpha(\mu),\,\,
\cos \theta=\mu,\,\,
-\sin \theta{{\partial }\over {\partial \theta}}=(1-\mu^2){{\partial
}\over {\partial \mu}}=D.
\end{eqnarray}
After some algebra, we get a single equation (Zaqarashvili et al. \cite{Zaqarashvili2009}):
%
%
\begin{eqnarray}\label{eq:general}
(\lambda D +s \mu)\left \{{1\over {s^2 -\epsilon
\lambda^2(1-\mu^2)}}\left [\lambda D -s \mu- {{\alpha^2}\over
{\lambda^2}}s^2\lambda(D +2\mu)\right ]\right \}{\hat u}_{\theta}-(\lambda^2 -\mu^2){\hat u}_{\theta}+s^2\alpha^2{\hat u}_{\theta} +
2\alpha^2\mu D{\hat u}_{\theta} +\mu s {\alpha^2\over
\lambda}(D+2\mu){\hat u}_{\theta}+\mu D(\alpha^2){\hat u}_{\theta}=0.
\end{eqnarray}
This is the general equation for wave dynamics in a shallow-water MHD system. It contains various kind of waves (magneto-gravity, Rossby, Kelvin).


Equation ~\ref{eq:general} can be rewritten as

$$
{{\partial }\over {\partial \mu}}\left (1-\mu^2 \right ){{\partial
{{\hat u}_{\theta}}}\over {\partial \mu}} -
\left [{{s+2\lambda}\over {\lambda^2-\alpha^2s^2}} \alpha^2 \lambda
{\epsilon} \mu + {{2\lambda^2{\epsilon} \mu}\over {s^2 -\epsilon
\lambda^2\left (1-\mu^2 \right )}} +{{2 \alpha s^2}\over
{\lambda^2-\alpha^2s^2}}{{\partial \alpha}\over {\partial \mu}}\right ] \left (1-\mu^2 \right ){{\partial {\hat u}_{\theta}}\over {\partial \mu}} +
\Big [ - {s^2\over {1-\mu^2}} + {\epsilon} \lambda^2 - {{s \lambda +
2\alpha^2s^2}\over {\lambda^2-\alpha^2s^2}} -
$$
\begin{equation}\label{eq:general1}
-{{{\epsilon} \lambda^2
+ 2{\epsilon} \alpha^2 s \lambda}\over {\lambda^2-\alpha^2s^2}}\mu^2
+ {{2{\epsilon}\lambda^3 s + 4{\epsilon} \alpha^2 s^2
\lambda^2}\over {(\lambda^2-\alpha^2s^2)[s^2 -\epsilon
\lambda^2\left (1-\mu^2 \right )]}}\mu^2-{{2 \alpha \epsilon \lambda^2 }\over
{\lambda^2-\alpha^2s^2}}{{\partial \alpha}\over {\partial
\mu}}\mu\left (1-\mu^2 \right ) - {{2 s^2 \alpha}\over
{\lambda^2-\alpha^2s^2}}{{\partial \alpha}\over {\partial \mu}}\mu
\Big ]{\hat u}_{\theta}=0.
\end{equation}
Now substitution of a new function
\begin{equation}
{\hat u}_{\theta}= u_{\theta1}\exp{1\over 2}
\int\left [{{s+2\lambda}\over {\lambda^2-\alpha^2s^2}} \alpha^2
\lambda {\epsilon} \mu + {{2\lambda^2{\epsilon} \mu}\over {s^2
-\epsilon \lambda^2\left (1-\mu^2 \right )}}+{{2 \alpha s^2}\over
{\lambda^2-\alpha^2s^2}}{{\partial \alpha}\over {\partial \mu}}\right ] d{\mu}
\end{equation}

gives

$$
{{\partial }\over {\partial \mu}}\left (1-\mu^2 \right ){{\partial
{u_{\theta1}}}\over {\partial \mu}} + \Big [ - {s^2\over
{1-\mu^2}} + {\epsilon} \lambda^2 - {{s \lambda + 2\alpha^2s^2}\over
{\lambda^2-\alpha^2s^2}} -{{{\epsilon} \lambda^2 + 2{\epsilon}
\alpha^2 s \lambda}\over {\lambda^2-\alpha^2s^2}}\mu^2 +
{{2{\epsilon} \lambda^3 s + 4{\epsilon} \alpha^2 s^2 \lambda^2}\over
{(\lambda^2-\alpha^2s^2)[s^2 -\epsilon \lambda^2\left (1-\mu^2
\right )]}}\mu^2 -{{s+2 \lambda}\over {\lambda^2-\alpha^2s^2}}\epsilon \alpha^2
\lambda \mu^2 -
$$
$$
-{{{2 \epsilon} \lambda^2}\over {s^2 -\epsilon
\lambda^2(1-\mu^2)}}\mu^2+{{{\epsilon} \lambda^2}\over {s^2
-\epsilon \lambda^2(1-\mu^2)}}\left (1-\mu^2 \right )+{1\over
2}{{s+2 \lambda}\over {\lambda^2-\alpha^2s^2}}\epsilon \alpha^2
\lambda\left (1-\mu^2 \right )-
{1\over 4}{{(s+2\lambda)^2}\over {(\lambda^2-\alpha^2s^2)^2}}
\epsilon^2 \alpha^4 \lambda^2 \mu^2 \left (1-\mu^2 \right ) -
$$
$$
-{{(s+2\lambda)\epsilon^2 \alpha^2 \lambda^3}\over
{(\lambda^2-\alpha^2s^2)[s^2 -\epsilon \lambda^2(1-\mu^2)]}}\mu^2
\left (1-\mu^2 \right ) -{{3\epsilon^2 \lambda^4}\over {[s^2 -\epsilon
\lambda^2(1-\mu^2)]^2}}\mu^2 \left (1-\mu^2 \right )-{{2\epsilon
\alpha \lambda^2 s^2}\over {(\lambda^2-\alpha^2s^2)[s^2 -\epsilon
\lambda^2(1-\mu^2)]}}\mu {{\partial \alpha}\over {\partial
\mu}}\left (1-\mu^2 \right )+
$$
\begin{equation}\label{eq:working}
+{{\epsilon \lambda -4s}\over {\lambda^2-\alpha^2s^2}}s \alpha
{{\partial \alpha}\over {\partial \mu}}  \mu \left (1-\mu^2 \right
)+ {{s^2 \lambda^2}\over {\left (\lambda^2-\alpha^2s^2\right
)^2}}\left ({{\partial \alpha}\over {\partial \mu}}\right )^2 \left
(1-\mu^2 \right )+{{\alpha s^2}\over
{\lambda^2-\alpha^2s^2}}{{\partial^2 \alpha}\over {\partial \mu^2}}
\left (1-\mu^2 \right )\Big ]u_{\theta1} =0.
\end{equation}

At this stage we should fix the magnetic field profile. We consider
two different cases of the magnetic field $B_{\phi}=B_0 \sqrt{1-\mu^2}$ and $B_{\phi}=
B_0 \mu \sqrt{1-\mu^2}$, where $B_0$ is constant parameter. Here, $B_{\phi}=B_0 \sqrt{1-\mu^2}$ means that the resulting magnetic field is at its
maximum at the equator and tends to zero at the poles. The magnetic field is opposite in the northern and southern hemispheres.
Then, $B_{\phi}=
B_0 \mu \sqrt{1-\mu^2} $ means that the magnetic field tends to zero at the solar equator and poles, reaching the maximum value at middle latitudes. This configuration is similar to the structure of the solar magnetic field, where the sunspots appear at middle latitudes (Gilman and Fox \cite{gilman97}). We now consider both cases separately.

\section{Uniform magnetic field ($B_{\phi}=B_0 \sqrt{1-\mu^2}$) }

The equation $B_{\phi}=B_0 \sqrt{1-\mu^2}$ implies $\alpha=const$ and Eq.~\ref{eq:working} is rewritten as
\begin{equation}\label{eq:constant}
{{\partial }\over {\partial \mu}}\left (1-\mu^2 \right ){{\partial {u_{\theta1}}}\over {\partial \mu}} + \left [ - {s^2\over {1-\mu^2}} + B_1 +
B_2\mu^2 + B_3\left (1-\mu^2 \right ) +B_4\mu^2\left (1-\mu^2 \right )\right ]u_{\theta1}=0,
\end{equation}
where
\begin{equation}\label{eq:constant1}
B_1={\epsilon} \lambda^2 - {{s \lambda + 2\alpha^2s^2}\over
{\lambda^2-\alpha^2s^2}} ,
\end{equation}
\begin{equation}\label{eq:constant2}
B_2= -{{{\epsilon} \lambda^2 + 2{\epsilon} \alpha^2 s \lambda}\over
{\lambda^2-\alpha^2s^2}} + {{2{\epsilon} \lambda^3 s + 4{\epsilon}
\alpha^2 s^2 \lambda^2}\over {(\lambda^2-\alpha^2s^2)[s^2 -\epsilon
\lambda^2(1-\mu^2)]}} -{{s+2 \lambda}\over
{\lambda^2-\alpha^2s^2}}\epsilon \alpha^2 \lambda - {{{2 \epsilon}
\lambda^2}\over {s^2 -\epsilon \lambda^2(1-\mu^2)}},
\end{equation}
\begin{equation}\label{eq:constant3}
B_3={{{\epsilon} \lambda^2}\over {s^2-\epsilon
\lambda^2(1-\mu^2)}}+{1\over 2}{{s+2 \lambda}\over
{\lambda^2-\alpha^2s^2}}\epsilon \alpha^2 \lambda,
\end{equation}
\begin{equation}\label{eq:constant4}
B_4= - {1\over 4}{{(s+2\lambda)^2}\over
{(\lambda^2-\alpha^2s^2)^2}}\epsilon^2 \alpha^4 \lambda^2 -
{{(s+2\lambda)\epsilon^2 \alpha^2 \lambda^3}\over
{(\lambda^2-\alpha^2s^2)[s^2 -\epsilon \lambda^2(1-\mu^2)]}} -
{{3\epsilon^2 \lambda^4}\over {[s^2 -\epsilon
\lambda^2(1-\mu^2)]^2}}.
\end{equation}

\begin{figure}[t]
\vspace*{1mm}
\begin{center}
\includegraphics[width=7cm]{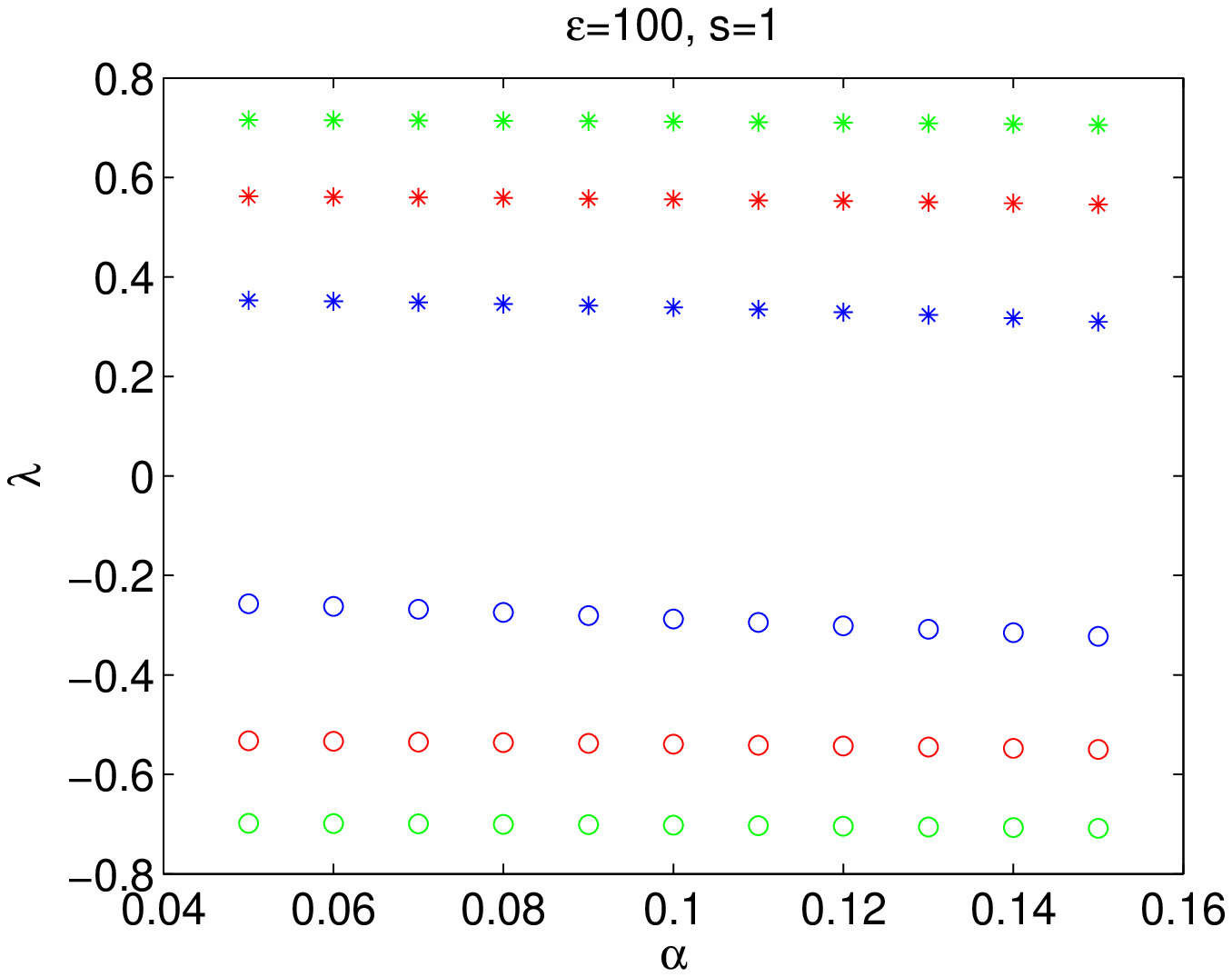}
\includegraphics[width=7cm]{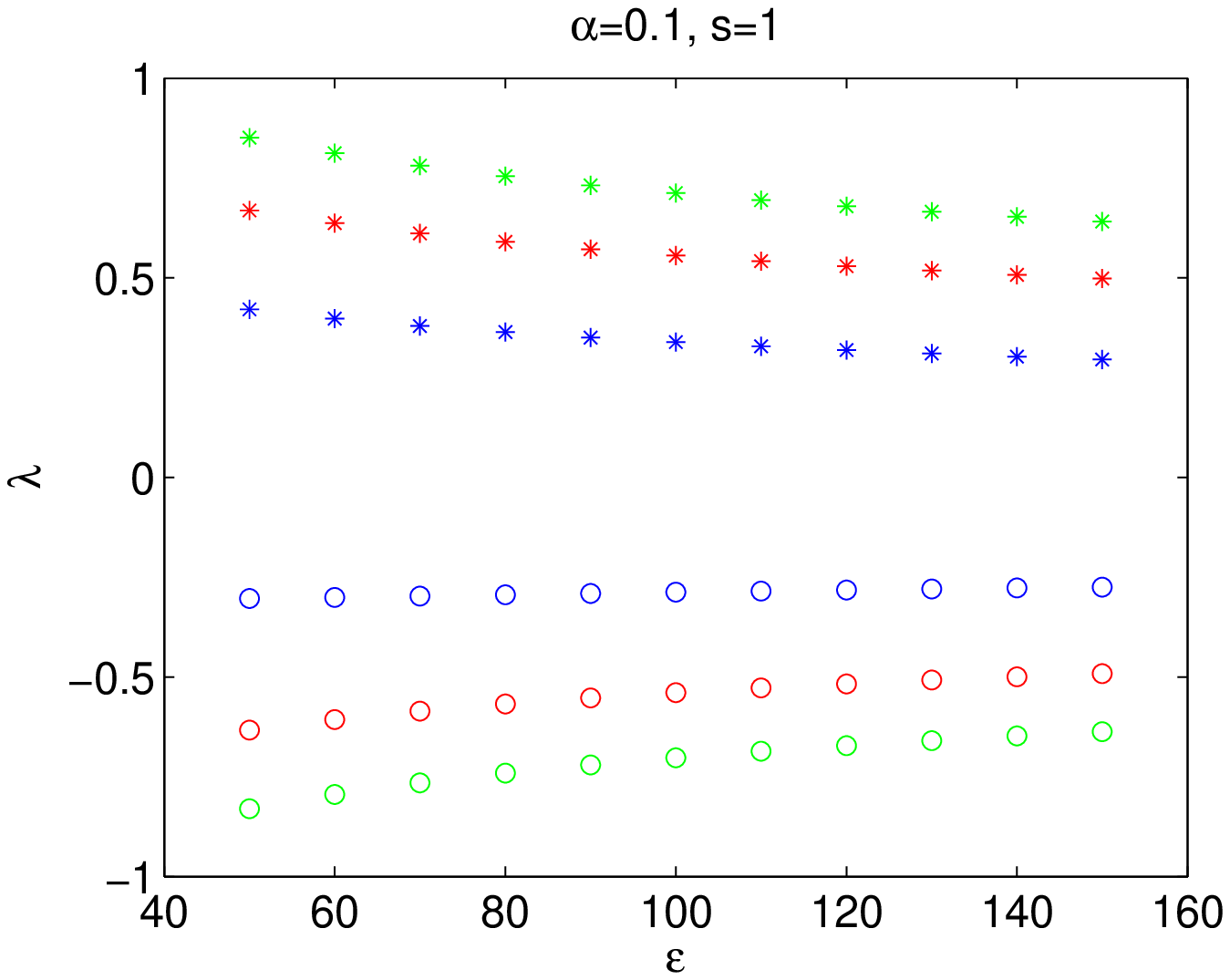}
\end{center}
\caption{Left: normalized frequency
${\lambda}={{\omega}/ {2\Omega_0}}$ vs ratio of Alfv\'en to surface
speed $\alpha=v_A/2\Omega_0R_0$. Here $s=1$ (toroidal wave number)
and $\epsilon={{4\Omega^2_0R^2_0}/{g H_0}}=100$ (ratio of surface to
internal gravity speed). The blue, red, and green asterisks (circles)
correspond to the positive (negative) solutions of the dispersion
relation Eq.~\ref{eq:w-dispersion} for the $\nu=0$, $\nu=1$, and
$\nu=2$ harmonics, respectively. Right: normalized
frequency ${\lambda}={{\omega}/{2\Omega_0}}$ vs $\epsilon$ for $s=1$
and $\alpha$=0.1.}
\end{figure}

We now consider the approximation of rapidly rotating stars or
a spherical shell with reduced gravity (such as in the overshoot
region):
\begin{equation}
{{\epsilon}}={{4\Omega^2_0R^2_0}\over {g H_0}}\gg 1.
\end{equation}
In this case Eq.~\ref{eq:constant} is a spheroidal wave equation,
but with a slightly complicated form (for example, the terms with
$\mu^4$ arise). An analytical method for the solution of such equations was proposed by Longuet-Higgins (\cite{Longuet-Higgins1965},\cite{Longuet-Higgins1968}) for the Earth's atmosphere. We closely follow this method.
The equations contain two different branches of solutions: high and low frequency ones. The higher frequency branch is represented by magnetic Poincar\'e waves and the lower frequency one by magnetic Rossby waves. We study both cases separately.

\subsection{High-frequency waves}

When $\epsilon \gg 1$, Eq.~\ref{eq:constant} is most easily
satisfied if $\lambda^2$, $\alpha^2$, and $\mu^2$ are small enough to compensate for the large $\epsilon$. Let us introduce a new independent variable $\eta = \epsilon^{1/4} \mu$ and consider that
$\lambda \sim \epsilon^{-1/4} L$ and $\alpha \sim \epsilon^{-1/2} L$, where L
is near unity.
Keeping only
the terms $\geq \epsilon^{1/4}$ in Eq.~\ref{eq:constant}, we get

\begin{equation}\label{eq:weber}
\left [{{\partial^2 }\over {\partial \eta^2}} + {A\over {\sqrt
{\epsilon}}} - \eta^2 \right ]u_{\theta1}=0.
\end{equation}
where
\begin{equation}\label{eq:A}
A= \epsilon \lambda^2 - {{s}\over {\lambda}} +
{1\over 2}{{\epsilon \alpha^2 s}\over {\lambda}}.
\end{equation}

\begin{figure}[t]
\vspace*{1mm}
\begin{center}
\includegraphics[width=7cm]{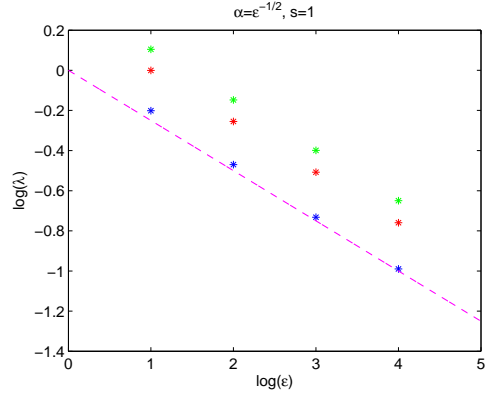}
\end{center}
\caption{Normalized frequency ${\lambda}={{\omega}/{2\Omega_0}}$ for four different values of $\epsilon$ (10, 100, 1000, 10000). Here $\epsilon$ is shown on a logarithmic scale. The blue, red, and green asterisks correspond to the positive solutions of the dispersion relation Eq.~\ref{eq:w-dispersion} for the $\nu=0$, $\nu=1$, and $\nu=2$ harmonics (with $s=1$), respectively. The dashed line corresponds to $\lambda=\epsilon^{-1/4}$ slope. Here the ratio of Alfv\'en to surface speed $\alpha=v_A/2\Omega_0R_0$ equals to $\epsilon^{-1/2}$ as assumed in the derivation of Eq.~\ref{eq:weber}.}
\end{figure}

This is the Weber (parabolic cylinder) equation, and the finite solutions at $\eta \rightarrow \pm \infty$ yield (Abramowitz and Stegun \cite{abramowitz})
\begin{equation}\label{eq:weber-dispersion}
{A\over {\sqrt {\epsilon}}}=2 \nu + 1 \,\, (\nu =0,1,2,...)
\end{equation}
and then
\begin{equation}\label{eq:weber-function}
u_{\theta1} \propto e^{-{1\over 2}\eta^2}H_{\nu}(\eta),
\end{equation}
where $H_{\nu}(\eta)$ is the Hermite polynomial $\nu$. The function $u_{\theta1}$ is exponentially small beyond the turning points

\begin{equation}
\eta=\sqrt {2 \nu + 1},
\end{equation}
that is to say,
\begin{equation}
\cos \theta= \epsilon^{-1/4} \sqrt {2 \nu + 1}.
\end{equation}
Due to the large $\epsilon$, the co-latitude $\theta$ satisfying
this condition is close to $90^0$, therefore the solution is
confined to the neighborhood of the equator.

\begin{figure}[t]
\vspace*{1mm}
\begin{center}
\includegraphics[width=6cm]{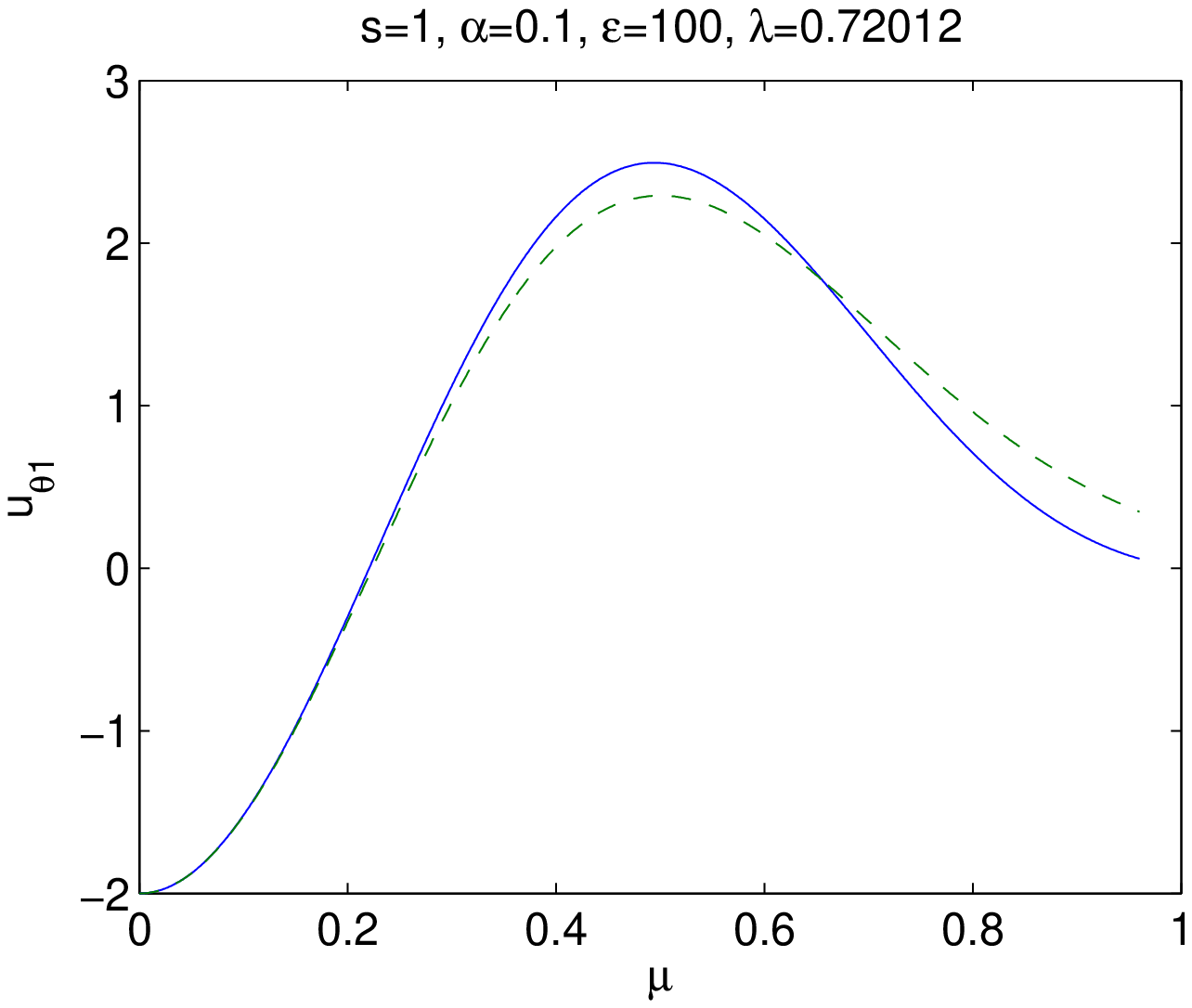}
\includegraphics[width=6cm]{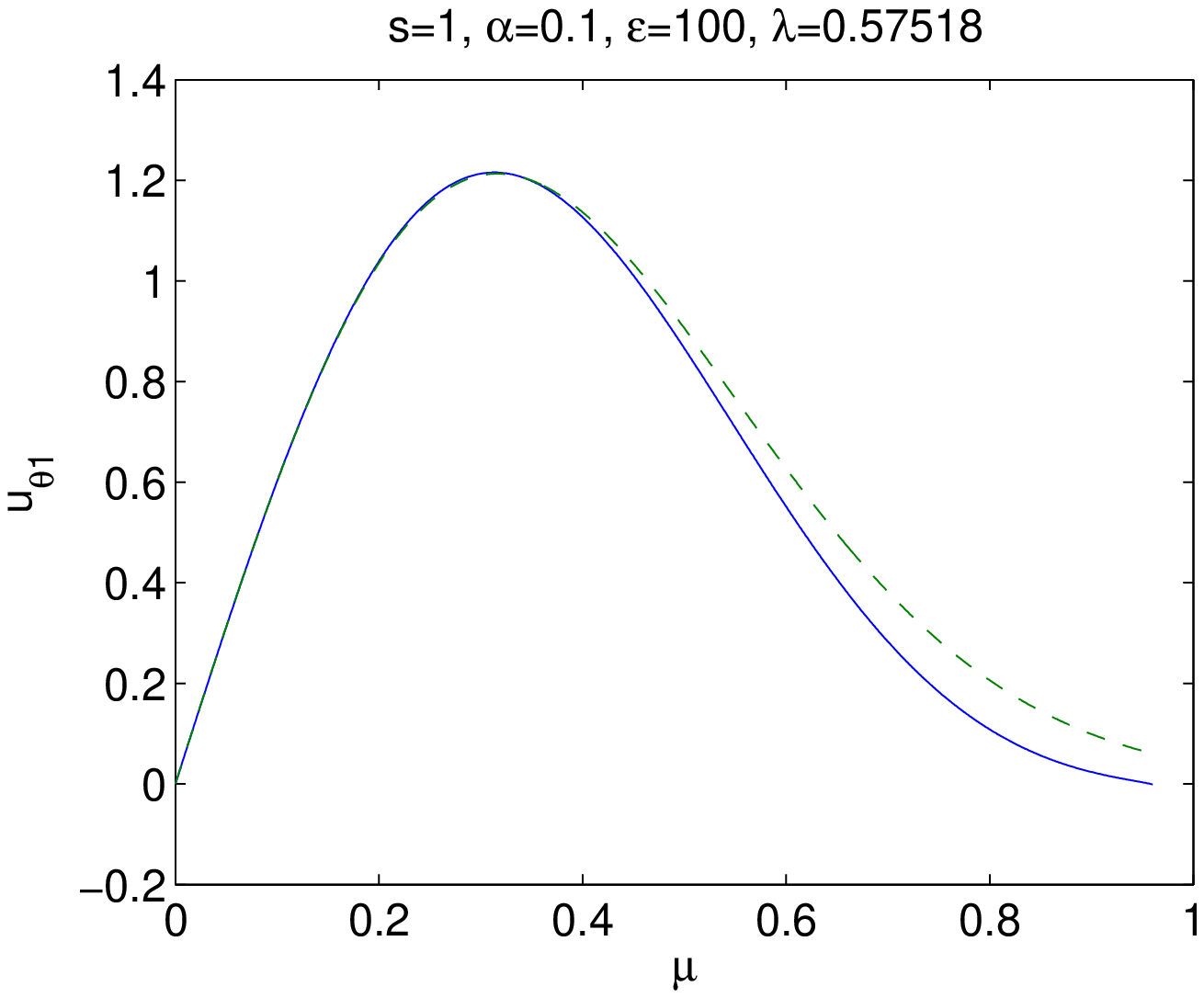}
\includegraphics[width=6cm]{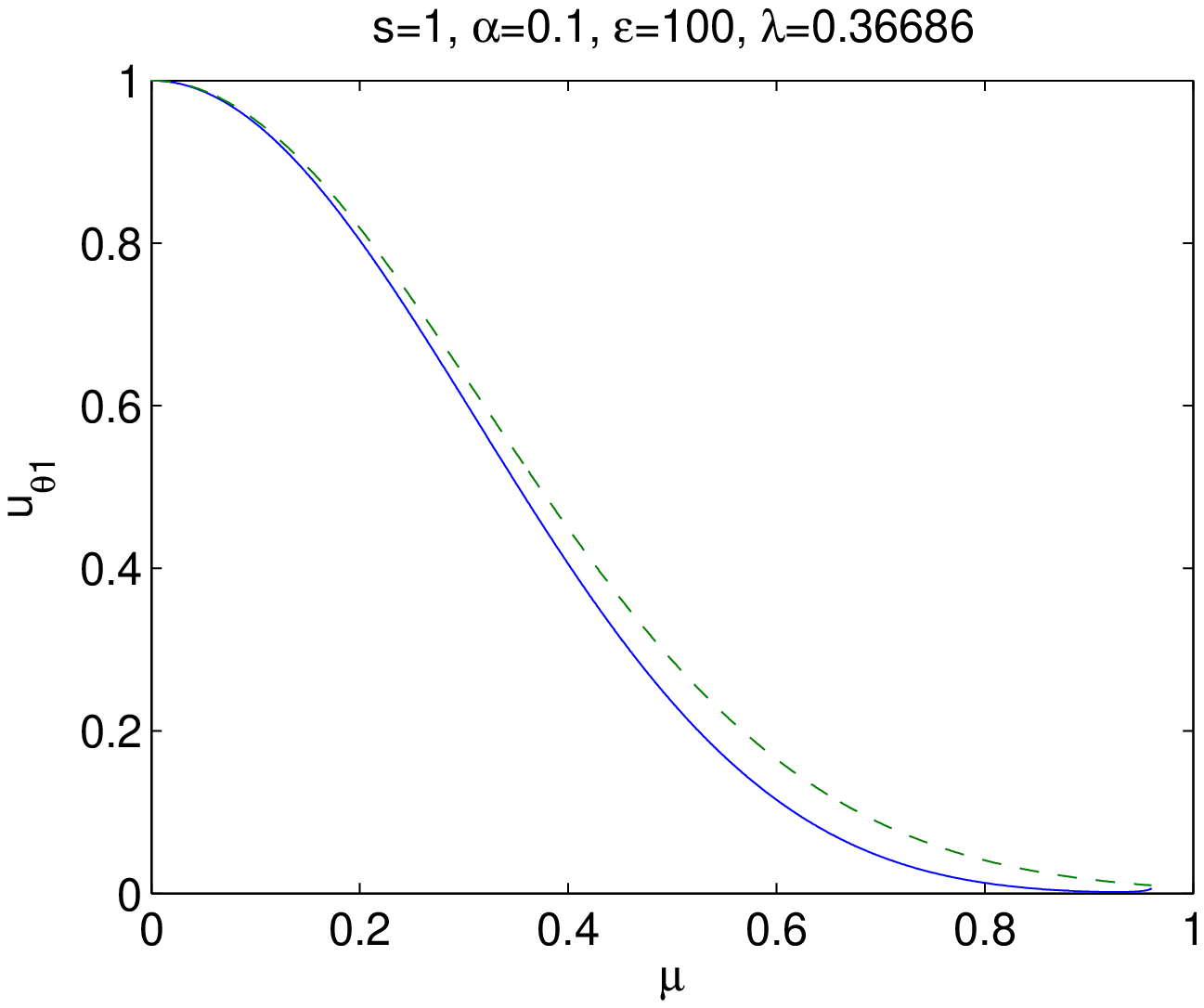}
\end{center}
\caption{Numerical solutions of Eq.~\ref{eq:constant} for three
different values of the normalized frequency ${\lambda}={{\omega}/
{2\Omega_0}}$ (solid lines). $\mu=\cos \theta$, where $\theta$ is
the co-latitude, hence $\mu=0$ corresponds to the equator and
$\mu=1$ corresponds to the northern pole. Upper, middle and lower
panels correspond to the harmonics with $\nu=2$ ($\lambda$=0.72012),
$\nu=1$ ($\lambda$=0.57518) and $\nu=0$ ($\lambda$=0.36686),
respectively. Dashed lines represent corresponding analytical
solutions from Eq.~\ref{eq:weber-function}. All solutions are
concentrated at low latitudes. We take $s=1$, $\epsilon=100$ and
$\alpha=0.1$ during the simulations.}
\end{figure}

Equation ~\ref{eq:weber-dispersion} yields the wave dispersion relation as

\begin{equation}\label{eq:w-dispersion}
\lambda^3 - {{2\nu +1}\over {\sqrt {\epsilon}}}\lambda -{s\over
{\epsilon}}\left (1 - {{\epsilon \alpha^2}\over 2} \right )=0.
\end{equation}
For non-magnetic case, this equation transforms into Eq. (8.11) of Longuet-Higgins (\cite{Longuet-Higgins1968}).

For high values of $\epsilon$, the solutions of this cubic equation can be given as
\begin{equation}\label{eq:w-dispersion1}
\lambda =\pm {{(2\nu +1)^{1/2}}\over {\epsilon^{1/4}}} + {s\over
{(4\nu +2)\epsilon^{1/2}}}\left (1 - {{\epsilon \alpha^2}\over 2}
\right ),
\end{equation}
 \begin{equation}\label{eq:w-dispersion2}
\lambda = - {s\over {(2\nu +1)\epsilon^{1/2}}}\left (1 - {{\epsilon
\alpha^2}\over 2} \right ).
\end{equation}

The first solution represents the high-frequency branch of shallow-water equations, i.e. magnetic Poincar\'e waves (or magneto-gravity waves).
For $\alpha=0$, expression (\ref{eq:w-dispersion1}) is the same as for the nonmagnetic case (Longuet-Higgins \cite{Longuet-Higgins1968}). The dispersion relation yields $\lambda \sim \epsilon^{-1/4}$, so it agrees with the approximation used in the derivation.
On the other hand, expression (\ref{eq:w-dispersion2}) represents the magnetic Rossby waves, but it does not agree with the approximations as it yields  $\lambda \sim \epsilon^{-1/2}$. Therefore, the second solution does not describe the wave dynamics correctly.

We solve the dispersion relation Eq.~\ref{eq:w-dispersion} numerically for different values of the ratio of Alfv\'en to surface speed $\alpha=v_A/2\Omega_0R_0$ and the ratio of surface to internal gravity speed $\epsilon={{4\Omega^2_0R^2_0}/{g
H_0}}$. Stability analysis of shallow-water MHD equations in the presence of latitudinal differential rotation shows that the $s=1$ harmonic is the most unstable one in all considered cases (Gilman and Fox \cite{gilman97}, Zaqarashvili et al. \cite{Zaqarashvili2010a},\cite{Zaqarashvili2010b}). Therefore, in what follows, we always consider the $s=1$ harmonics.

The normalized frequency ${\lambda}={{\omega}/
{2\Omega_0}}$ vs $\alpha$ and $\epsilon$ is shown on Fig. 1. The left hand panel shows
the dependence of ${\lambda}$ on $\alpha$ for $s=1$ and
$\epsilon=100$.
We see that the frequency depends only slightly on $\alpha$, i.e. on
magnetic field strength. The right hand panel shows the dependence
of ${\lambda}$ on $\epsilon$ for $s=1$ and $\alpha=0.1$. The
dependence of the wave frequency on $\epsilon$ is more pronounced;
higher $\epsilon$ yields smaller ${\lambda}$. The figure displays
only those two solutions of the dispersion relation
Eq.~\ref{eq:w-dispersion} that satisfy the condition $\lambda \sim
\epsilon^{-1/4}$. The third solution, which has lower frequency and
does not satisfy $\lambda \sim \epsilon^{-1/4}$, is not shown here.

The normalized frequency ${\lambda}$ for four
different values of $\epsilon$ (10, 100, 1000, 10000) is shown on Fig. 2. The figure
shows that the increase of $\epsilon$ leads to a significant
decrease in the normalized frequency. For example, $\epsilon=10$
yields $T_{\mathrm {osc}} \approx 0.8 \, T_{\mathrm {rot}}$, while $\epsilon=10000$ yields
$T_{\mathrm {osc}} \approx 5\, T_{\mathrm {rot}}$. Here $T_{\mathrm {osc}}$ is the oscillation
period of $s=1$, $\nu=0$ harmonics, and $T_{\mathrm {rot}}$ is the stellar
rotational period. This means that faster rotation leads to a longer
oscillation period than the rotational period.

To check the analytical solutions, we solved Eq.~\ref{eq:constant} numerically for several frequencies, which were
found from the dispersion relation Eq.~\ref{eq:w-dispersion}. We
searched the bounded solutions of Eq.~\ref{eq:constant} varying
$\lambda$ close to its analytical counterpart. The
numerical solutions of Eq.~\ref{eq:constant} for three different
values of ${\lambda}$ are shown on Fig. 3. We see that all solutions are
bounded and concentrated at low latitudes as suggested by the
analytical solution. The frequencies for which the bounded solutions
are calculated are in good coincidence with those obtained from the
analytical dispersion relation Eq.~\ref{eq:w-dispersion}. Therefore, the wave modes with frequency $\lambda \sim \epsilon^{-1/4}$ are concentrated at low latitudes. These modes are magnetic Poincar\'e waves (Zaqarashvili et al. \cite{Zaqarashvili2007}).

\subsection{Low-frequency waves}

We now concentrate on lower frequency waves by considering
$\lambda=\epsilon^{-1/2}L$ and $\alpha=\epsilon^{-1/2} L$, where $L$
is close to unity. Then $\epsilon \lambda^2 \sim 1$, which means that $s^2-\epsilon
\lambda^2(1-\mu^2)$ tends to zero for $s=1$ and small $\mu$. As a result, some terms
in Eq.~\ref{eq:constant} tend to infinity.
Therefore, Eq.~\ref{eq:constant} is most easily satisfied if
$1-\mu^2$ is small instead of $\mu^2$. This means that the solutions are located near
the poles not near the equator. We
again follow the steps of Longuet-Higgins. Returning back to the
variable $\theta$, Eq.~\ref{eq:constant} is written
\begin{equation}\label{eq:low-weber}
\sin \theta{{\partial }\over {\partial \theta}} \left (\sin \theta{\partial {u_{\theta1}}\over {\partial \theta}}\right ) + [ - s^2 + B_1\sin^2 \theta +B_2\cos^2 \theta \sin^2 \theta +B_3\sin^4 \theta + B_4\sin^4 \theta \cos^2 \theta ]u_{\theta1}=0.
\end{equation}
For small $\theta$ we have
\begin{equation}\label{eq:low-weber-theta}
{{\partial^2 {u_{\theta1}}}\over {\partial \theta^2}} + {1\over
\theta}{{\partial {u_{\theta1}}}\over {\partial \theta}} + \left [ -
{s^2\over \theta^2} + B_1 + B_2(1 - \theta^2) + B_3\theta^2
+B_4\theta^2(1-\theta^2)\right ]u_{\theta1}=0.
\end{equation}

\begin{figure}[t]
\vspace*{1mm}
\begin{center}
\includegraphics[width=7cm]{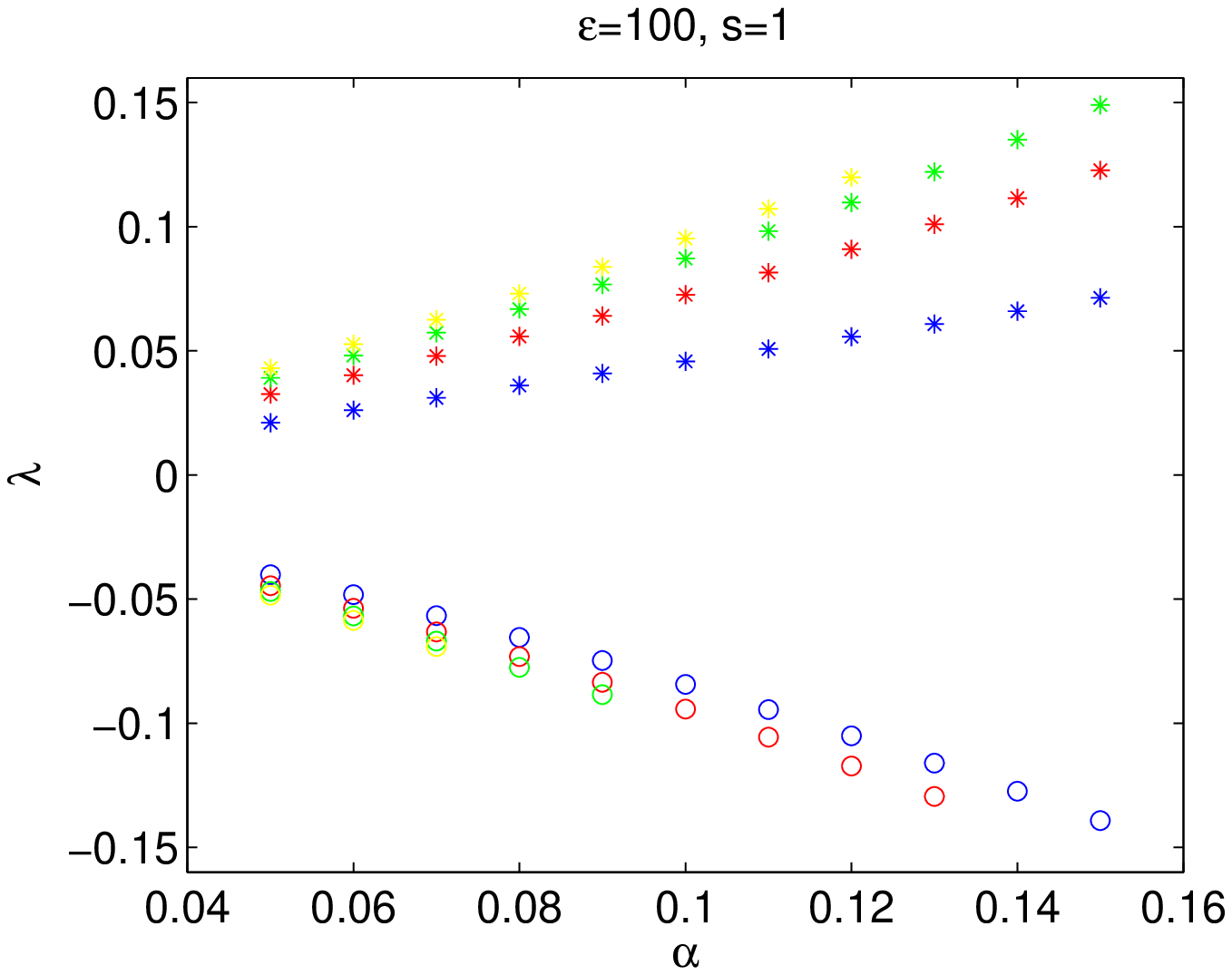}
\includegraphics[width=7cm]{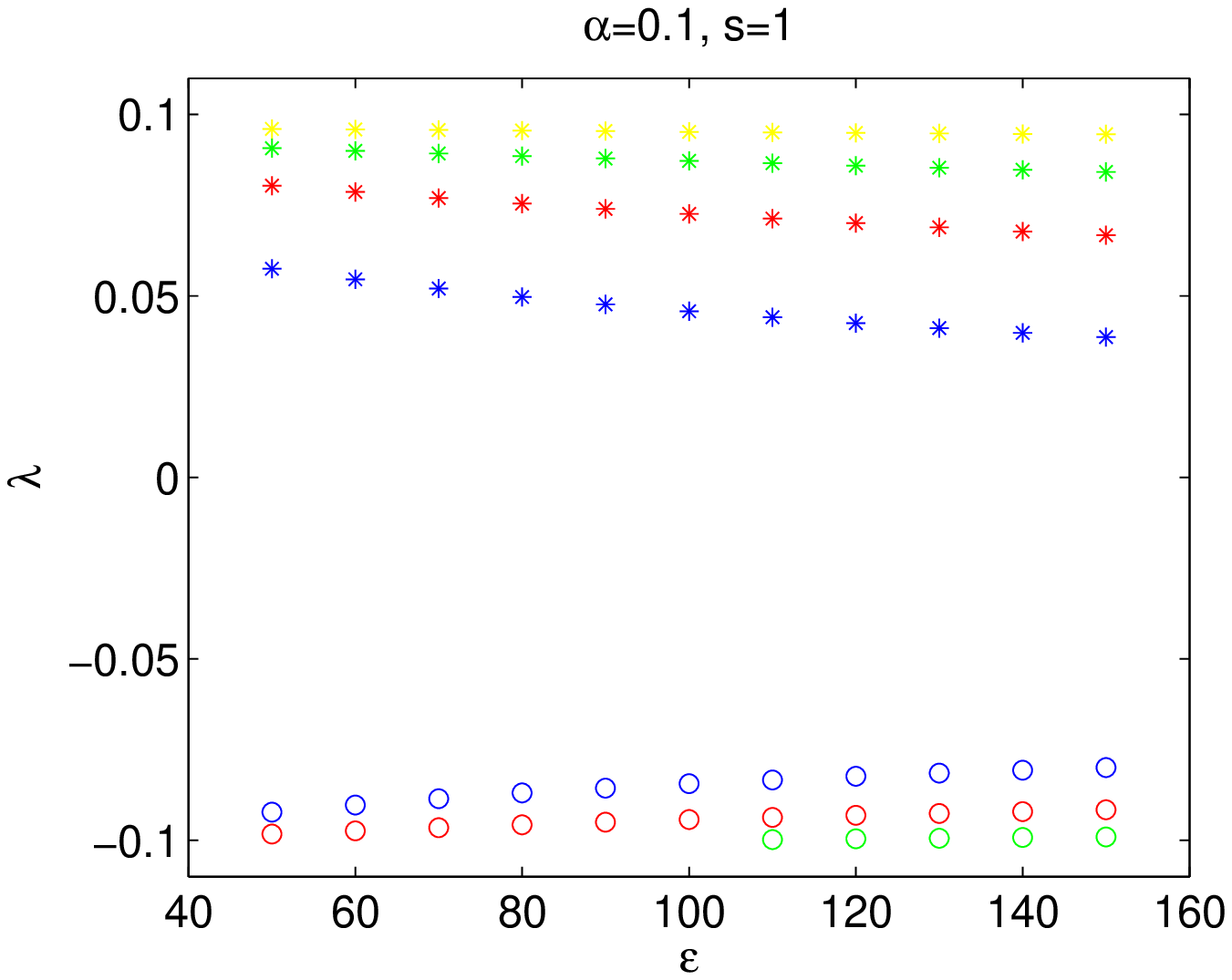}
\end{center}
\caption{Left: normalized frequency
${\lambda}={{\omega}/ {2\Omega_0}}$ vs ratio of Alfv\'en to surface
speed $\alpha=v_A/2\Omega_0R_0$. Here $s=1$ (toroidal wave number)
and $\epsilon={{4\Omega^2_0R^2_0}/{g H_0}}=100$ (ratio of surface to
internal gravity speed). The blue, red, green, and yellow asterisks
(circles) correspond to the positive (negative) solutions of the
dispersion relation Eq.~\ref{eq:const-disp} for the $n=1$, $n=2$,
$n=3$, and $n=4$ harmonics, respectively. Right:
normalized frequency ${\lambda}={{\omega}/{2\Omega_0}}$ vs
$\epsilon$ for $s=1$ and $\alpha=0.1$.}
\end{figure}

In order to estimate different terms in this equation, let us consider
$\theta = \epsilon^{-1/4} \eta$, where $\eta$ is near unity.
Keeping only the terms $\geq \epsilon^{1/2}$ we get from Eq.~\ref{eq:low-weber-theta}
\begin{equation}\label{eq:low-weber-theta1}
{{\partial^2 {u_{\theta1}}}\over {\partial \theta^2}} +
{1\over \theta}{{\partial {u_{\theta1}}}\over {\partial
\theta}} + \left [ - {s^2\over \theta^2} + A_1 - A_2\theta^2\right
]u_{\theta1}=0,
\end{equation}
where
\begin{equation}
A_1= -{{s \lambda}\over {\lambda^2-\alpha^2s^2}}-{{{\epsilon}
\lambda^2 + 3{\epsilon} \alpha^2 s \lambda}\over
{\lambda^2-\alpha^2s^2}} + {{2{\epsilon} \lambda^3}\over
{(\lambda^2-\alpha^2s^2)s}},
\end{equation}
\begin{equation}
A_2= -{{{\epsilon} \lambda^2}\over {\lambda^2-\alpha^2s^2}} +{1\over
4}{{\epsilon^2 \alpha^4 \lambda^2 s^2}\over
{(\lambda^2-\alpha^2s^2)^2}}.
\end{equation}

\begin{figure}
\vspace*{1mm}
\begin{center}
\includegraphics[width=7cm]{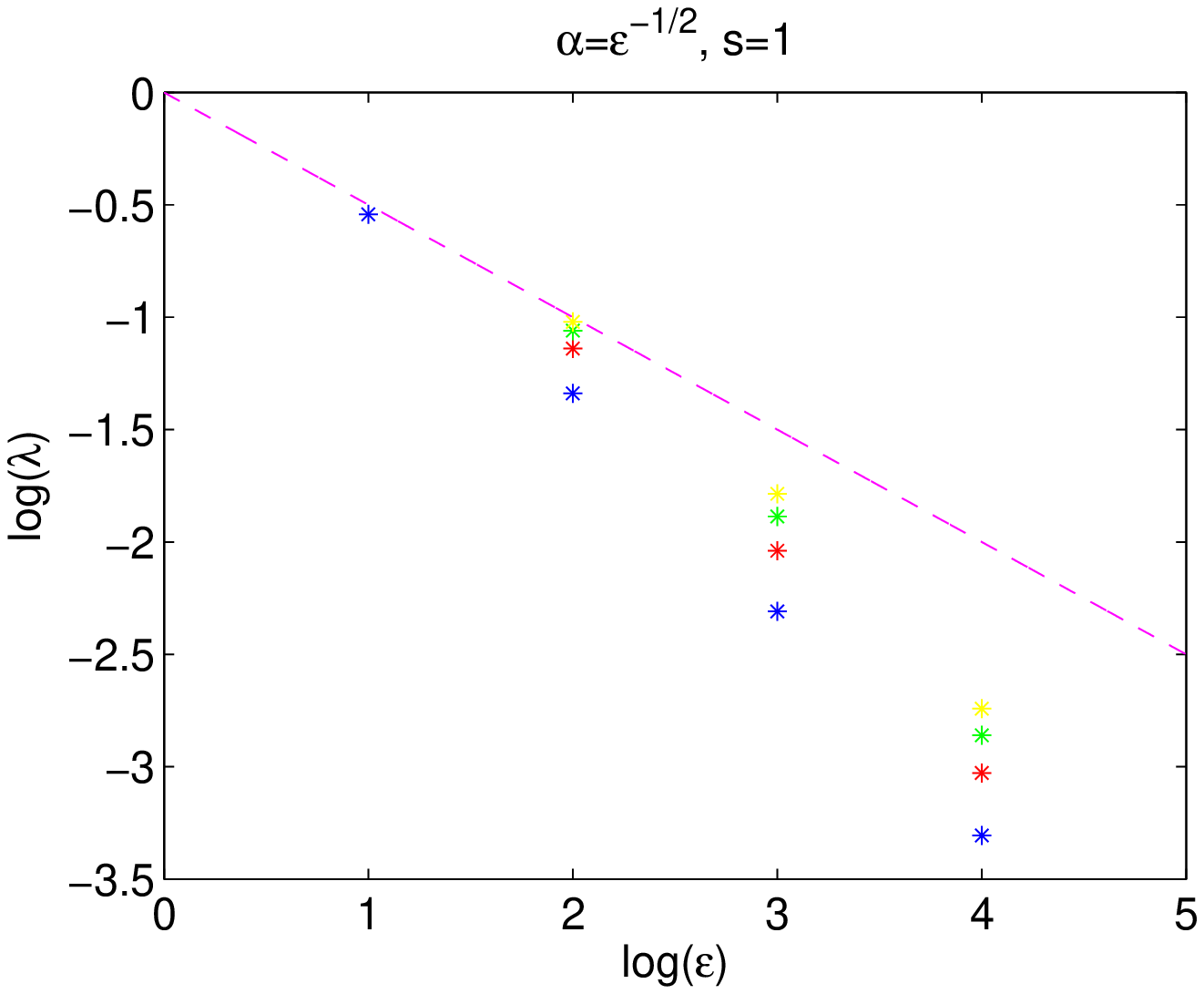}
\includegraphics[width=7cm]{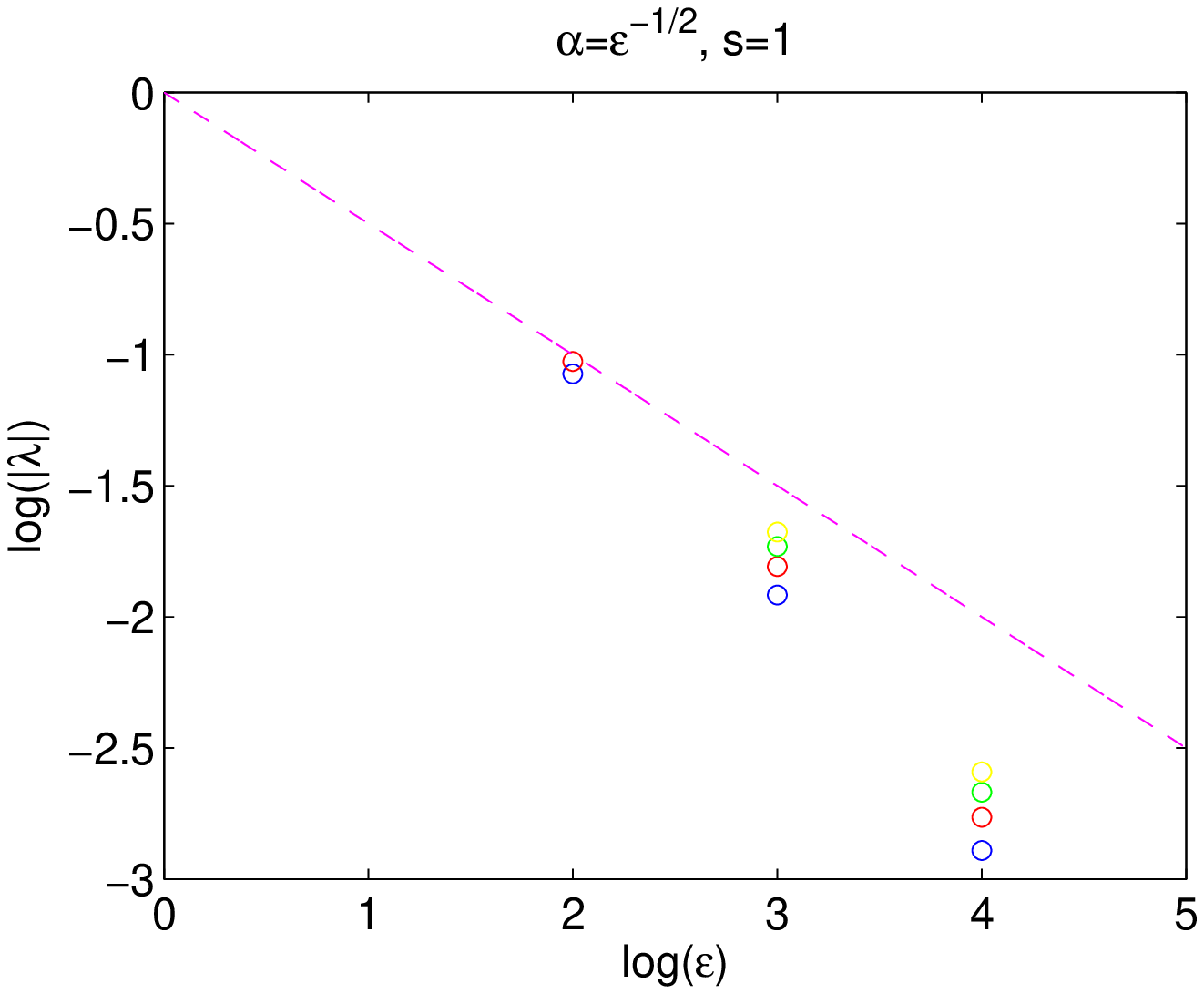}
\end{center}
\caption{Normalized frequency ${\lambda}={{\omega}/ {2\Omega_0}}$
for four different values of $\epsilon$ (10, 100, 1000, 10000). Here ${\lambda}$ and $\epsilon$ are shown on a
logarithmic scale. The blue,
red, green, and yellow asterisks (circles) correspond to the positive
(negative) solutions of dispersion relation Eq.~\ref{eq:const-disp}
for the $n=1$, $n=2$, $n=3$, and $n=4$ harmonics, respectively.
Dashed lines correspond to $\lambda=\epsilon^{-1/2}$ slope. Here the
ratio of Alfv\'en to surface speed $\alpha=v_A/2\Omega_0R_0$ equals
$\epsilon^{-1/2}$ as assumed in the derivation of
Eq.~\ref{eq:low-weber-theta}.}
\end{figure}

Now if we consider mode frequencies in the range
\begin{equation}
\lambda^2 < \alpha^2s^2 \left (1+ {{\epsilon \alpha^2}\over 4} \right ),
\end{equation}
but $\lambda \not= \pm \alpha s$, then $A_2$ is positive, and we can
define a new variable $x= \sqrt {A_2} \theta^2$, which gives the
equation
\begin{equation}\label{eq:laguerre-constant}
x{{\partial^2 {u_{\theta1}}}\over {\partial x^2}}
+{{\partial {u_{\theta1}}}\over {\partial x}} + \left
[-{x\over 4}+{{2n+s+1}\over 2} - {{s^2}\over {4 x}} \right ]u_{\theta1}=0,
\end{equation}
where $n$ is a nonnegative integer, and
\begin{equation}\label{eq:laguerre-disp}
2(2n +s +1)={A_1\over {\sqrt {A_2}}}.
\end{equation}
This equation is self-adjoint and its solutions are sometimes called
Laguerre functions $\psi^s_n(x)$ (see for example, Arfken and Weber \cite{Arfken2005}). These functions are bounded at $x=0$ and $x=\infty$ and are
linked to the associated Laguerre polynomials as
\begin{equation}\label{eq:laguerre-sol}
u_{\theta1}=u_{\theta1}(0)x^{{{1}\over 2}s}e^{-{1\over 2}x}L^{s}_n(x),
\end{equation}
where
\begin{equation}
L^{s}_n(x)=\sum^{n}_{k=0} {{(s+n)!}\over {(s+k)!(n-k)!k!}}(-x)^k,
\end{equation}
so that $s>-1$.

\begin{figure*}
\vspace*{1mm}
\begin{center}
\includegraphics[width=7cm]{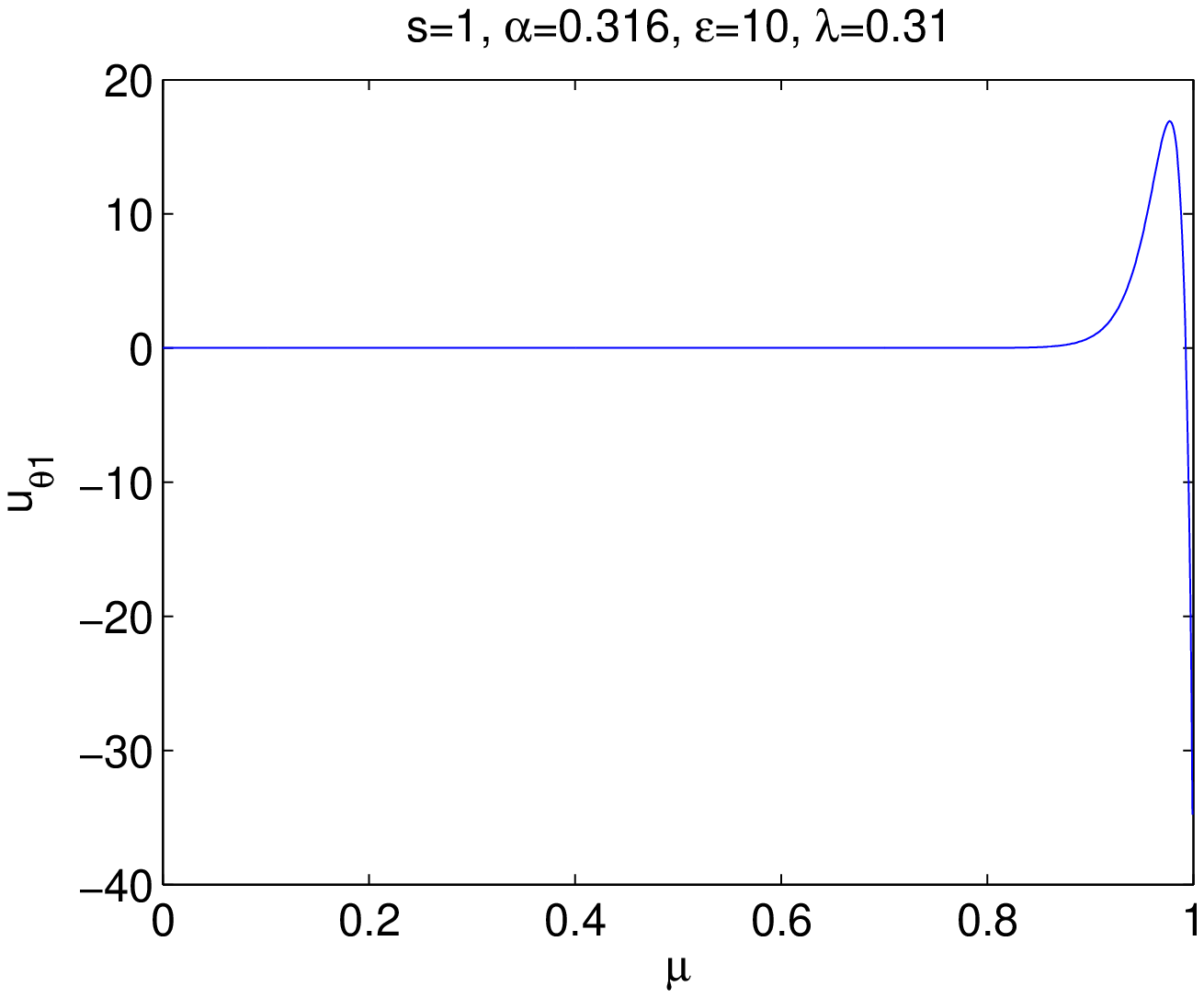}
\includegraphics[width=7cm]{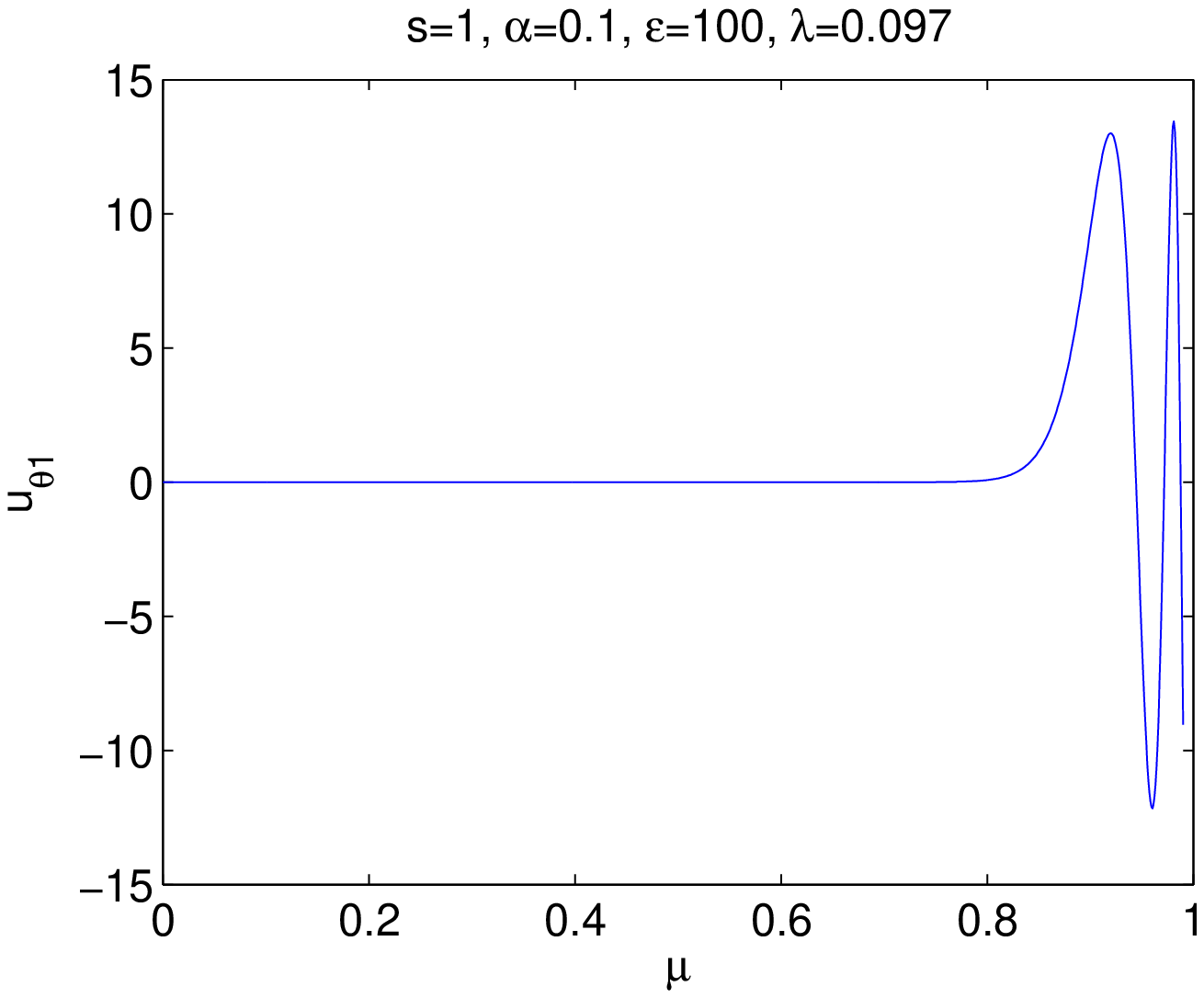}
\includegraphics[width=7cm]{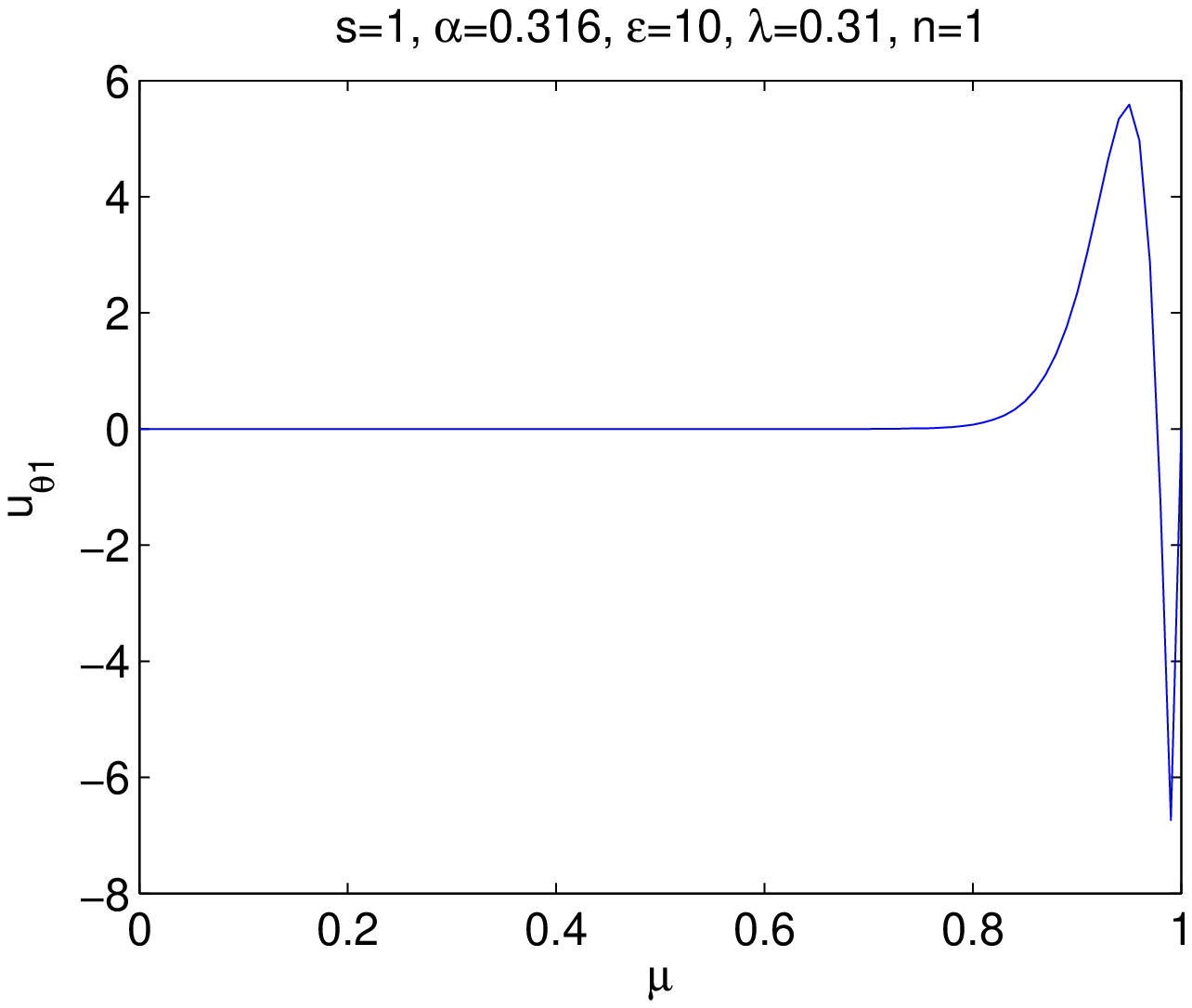}
\includegraphics[width=7cm]{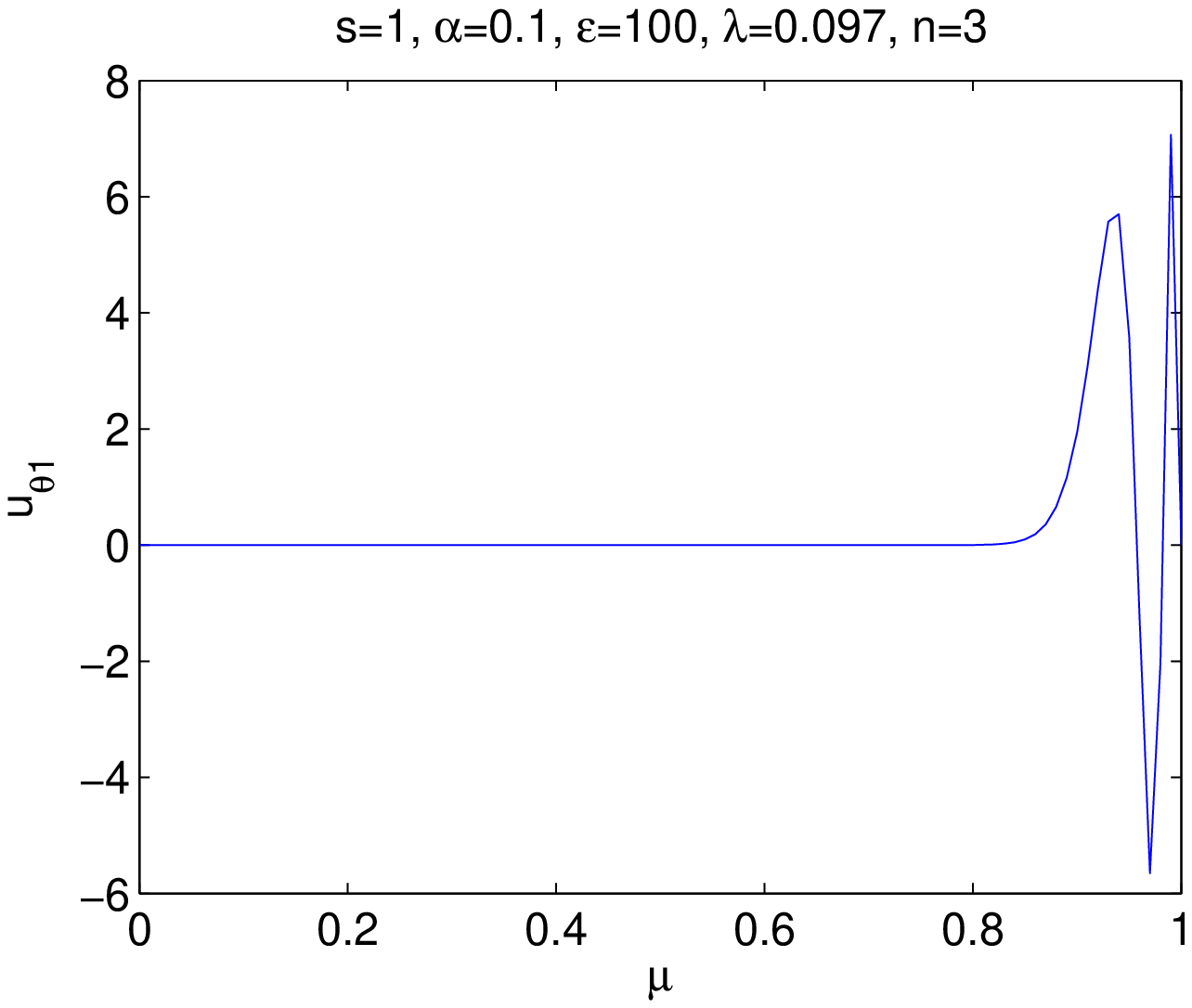}
\end{center}
\caption{Upper panels: Numerical solutions of Eq.~\ref{eq:constant} for two different values of $\epsilon$ and
corresponding normalized frequency ${\lambda}={{\omega}/
{2\Omega_0}}$.  $\mu=\cos \theta$, where $\theta$ is co-latitude,
hence $\mu=0$ corresponds to the equator and $\mu=1$ corresponds to
the northern pole. We take $s=1$ and $\alpha=\sqrt{\epsilon}$ during
the simulations. Lower panels: Corresponding analytical
solutions plotted from Eq.~\ref{eq:laguerre-sol}; left panel shows
the solution with $n=1$ and the right panel shows the solution with
$n=3$. }
\end{figure*}

Then Eq.~\ref{eq:laguerre-disp} gives the wave dispersion relation as
\begin{equation}\label{eq:const-disp}
{{2\epsilon}\over s}\lambda^2 - \epsilon \lambda -s \left (1 +
3{{\epsilon \alpha^2}}  \right )=\pm 2({{2n+s+1}})\sqrt {- \epsilon
\lambda^2 + \epsilon \alpha^2 s^2 + {1\over 4}{\epsilon^2 \alpha^4
s^2}}.
\end{equation}
The sign of $\lambda^2 -\alpha^2 s^2$ and $\lambda$ define the sign
of the righthand side of Eq.~\ref{eq:const-disp}:
$${\lambda\over {\lambda^2 - \alpha^2 s^2}}>0 \,\, \left ({\lambda\over {\lambda^2 - \alpha^2 s^2}}<0 \right ),$$
yields a + (-) sign. We solve the dispersion relation Eq.~\ref{eq:const-disp} numerically for different values of $\alpha$ and $\epsilon$.

The normalized frequency ${\lambda}$ vs $\alpha$ and
$\epsilon$ is shown on Fig. 4. The left hand panel shows the dependence of ${\lambda}$
on $\alpha$ for $s=1$ and $\epsilon=100$.
We see that unlike the previous case here the frequency
significantly depends on $\alpha$, i.e., on magnetic field strength:
higher $\alpha$ yields larger ${\lambda}$. The right hand panel
shows the dependence of ${\lambda}$ on $\epsilon$ for $s=1$ and
$\alpha=0.1$. The dependence of wave frequency on $\epsilon$ is less
pronounced here.

The normalized frequency ${\lambda}$ for four different values of $\epsilon$ (10, 100, 1000, 10000) is displaced on Fig. 5 .
The figure shows that the increase of $\epsilon$ leads to a significant decrease in the normalized frequency. For example, $\epsilon=10$ yields  $T_{\mathrm {osc}} \approx 1.75 T_{\mathrm {rot}}$, while $\epsilon=10000$ yields $T_{\mathrm {osc}} \approx 1000 T_{\mathrm {rot}}$. Here $T_{\mathrm {osc}}$ is the oscillation period of the $s=1$, $n=1$ harmonics and $T_{\mathrm {rot}}$ is the stellar rotational period. This means that faster rotation leads to a longer oscillation period with respect to the rotational period. However, the increase of $\epsilon$ leads to the deviation of $\lambda$ from $\epsilon^{-1/2}$, while the analytical solution is obtained in the approximation when $\lambda$ is the same order of magnitude as $\epsilon^{-1/2}$.

To check the analytical solutions, we solved Eq.~\ref{eq:constant} numerically for several frequencies, which were
found from the dispersion relation Eq.~\ref{eq:const-disp}. We
searched the solutions of Eq.~\ref{eq:constant} fixing $\lambda$
close to its analytical counterpart. The numerical
solutions of Eq.~\ref{eq:constant} for two different values of
$\epsilon$ and corresponding normalized frequency ${\lambda}$ is displaced on Fig. 6 (upper
panels). We see that all solutions are concentrated near the polar
regions, as suggested by the analytical solution. The
frequencies for which the solutions are calculated are in good
coincidence with those obtained from the analytical dispersion
relation Eq.~\ref{eq:const-disp}. The lower panels of this figure
display the corresponding analytical solutions plotted from Eq.~\ref{eq:laguerre-sol}. The figure shows that the analytical and
numerical solutions show the same latitudinal behavior.

Therefore, the wave modes with low frequency, i.e. $\lambda \sim \epsilon^{-1/2}$, are concentrated near the poles. These modes are magnetic Rossby waves (Zaqarashvili et al. \cite{Zaqarashvili2007}).

\section{Non-uniform magnetic field ($B_{\phi}=B_0 \mu \sqrt{1-\mu^2}$) }

In this case, Eq. ~\ref{eq:working} is rewritten as
\begin{equation}\label{eq:equation-mu}
{{\partial }\over {\partial \mu}}\left (1-\mu^2 \right ){{\partial
{u_{\theta1}}}\over {\partial \mu}} + \left [ - {s^2\over {1-\mu^2}}
+ B_1 + B_2\mu^2 + B_3\left (1-\mu^2 \right ) +B_4\mu^2\left
(1-\mu^2 \right )+B_5 \mu^4+B_6\mu^4\left (1-\mu^2 \right
)+B_7\mu^6\left (1-\mu^2 \right )\right ]u_{\theta1}=0,
\end{equation}
where
\begin{equation}
B_1={\epsilon} \lambda^2 - {{s \lambda}\over {\lambda^2-s^2\alpha_0^2\mu^2}} ,
\end{equation}
\begin{equation}
B_2= -{{{\epsilon} \lambda^2}\over {\lambda^2-s^2\alpha_0^2\mu^2}} -
{{2\alpha_0^2s^2}\over {\lambda^2-s^2\alpha_0^2\mu^2}}+
{{2{\epsilon} \lambda^3 s }\over {(\lambda^2-s^2\alpha_0^2\mu^2)[s^2
-\epsilon \lambda^2(1-\mu^2)]}}- {{{2 \epsilon} \lambda^2}\over {s^2
-\epsilon \lambda^2(1-\mu^2)}},
\end{equation}
\begin{equation}
B_3={{{\epsilon} \lambda^2}\over {s^2 -\epsilon
\lambda^2(1-\mu^2)}}+{{\lambda^2 \alpha_0^2 s^2}\over
{(\lambda^2-s^2\alpha_0^2\mu^2)^2}},
\end{equation}
\begin{equation}
B_4= - {{3\epsilon^2 \lambda^4}\over {[s^2 -\epsilon
\lambda^2(1-\mu^2)]^2}} -{{2\epsilon \alpha_0^2 \lambda^2 s^2}\over
{(\lambda^2-s^2\alpha_0^2\mu^2)[s^2 -\epsilon
\lambda^2(1-\mu^2)]}}+{{\epsilon \lambda -4s}\over
{\lambda^2-s^2\alpha_0^2\mu^2}}s \alpha_0^2+{1\over 2}{{s+2
\lambda}\over {\lambda^2-s^2\alpha_0^2\mu^2}}\epsilon \alpha_0^2
\lambda,
\end{equation}
\begin{equation}
B_5=-{{2{\epsilon} \alpha_0^2s \lambda}\over
{\lambda^2-s^2\alpha_0^2\mu^2}}+ {{4{\epsilon} \alpha_0^2 s^2
\lambda^2}\over {(\lambda^2-s^2\alpha_0^2\mu^2)[s^2 -\epsilon
\lambda^2(1-\mu^2)]}} -{{s+2 \lambda}\over
{\lambda^2-s^2\alpha_0^2\mu^2}}\epsilon \alpha_0^2 \lambda,
\end{equation}
\begin{equation}
B_6= -{{(s+2\lambda)\epsilon^2 \alpha_0^2 \lambda^3}\over
{(\lambda^2-s^2\alpha_0^2\mu^2)[s^2 -\epsilon \lambda^2(1-\mu^2)]}},
\end{equation}
\begin{equation}
B_7= -{1\over 4}{{(s+2\lambda)^2}\over
{(\lambda^2-s^2\alpha_0^2\mu^2)^2}} \epsilon^2 \alpha_0^4\lambda^2.
\end{equation}

We follow the steps of the previous section and
introduce a new independent variable $\eta = \epsilon^{1/4} \mu$, and also
consider $\lambda=\epsilon^{-1/4}L$ and $\alpha=\epsilon^{-1/2}
L$, where L is near unity. Keeping only the terms $\geq \epsilon^{1/4}$ in Eq.~\ref{eq:equation-mu} we get

%
%

\begin{equation}
\left [{{\partial^2 }\over {\partial \eta^2}} + {A\over {\sqrt
{\epsilon}}} - \eta^2 \right ]u_{\theta1}=0,
\end{equation}
where
\begin{equation}
A= \epsilon \lambda^2 - {{s}\over {\lambda}} +
{{\alpha_0^2 s^2}\over {\lambda^2}}.
\end{equation}

This equation and the corresponding dispersion relation was already
studied in detail by Zaqarashvili et al. (\cite{Zaqarashvili2009}), so we do not repeat the
calculations here. The solutions are concentrated near the equator
as in the previous section. The resulting dispersion relation contains magnetic Poincar\'e and magnetic Rossby waves. The influence of the magnetic field on Poincar\'e waves is even less pronounced in this profile than in the previous section (compare Eq.~\ref{eq:w-dispersion1} of the previous section to Eq.~11 in Zaqarashvili et al. \cite{Zaqarashvili2009}). The reason is that the magnetic field with this profile tends to zero near the equatorial regions, where the waves are concentrated. Therefore, it has almost no influence on the wave dynamics.

However, we now concentrate on lower frequency waves
considering $\lambda=\epsilon^{-1/2}L$ and $\alpha_0=\epsilon^{-1/2}
L$, where $L$ is near unity. Then, some terms
in Eq.~\ref{eq:constant} tend to infinity for $s=1$ and small $\mu$. Therefore, Eq.~\ref{eq:constant} is most
easily satisfied if $1-\mu^2$ is small instead of $\mu^2$. This
means that the solutions are located near the poles in this case.
Returning back to the variable $\theta$ Eq.~\ref{eq:equation-mu}
is written

$$
\sin \theta{{\partial }\over {\partial \theta}} \left (\sin
\theta{\partial {u_{\theta1}}\over {\partial \theta}}\right
) + [ - s^2 + B_1\sin^2 \theta + B_2\cos^2 \theta \sin^2 \theta +
B_3\sin^4 \theta +B_4\sin^4 \theta \cos^2 \theta + B_5\sin^2 \theta
\cos^4 \theta +B_6\sin^4 \theta \cos^4 \theta+
$$
\begin{equation}\label{eq:low-weber-mu}
+B_7\sin^4 \theta \cos^6 \theta]u_{\theta1}=0.
\end{equation}

For small $\theta$ we have
\begin{equation}\label{eq:low-weber-mu-theta}
{{\partial^2 {u_{\theta1}}}\over {\partial \theta^2}} +
{1\over \theta}{{\partial {u_{\theta1}}}\over {\partial
\theta}} + \left [ - {s^2\over \theta^2} + B_1 + B_2(1 - \theta^2) +
B_3\theta^2 + B_4\theta^2(1-\theta^2)+
B_5(1-\theta^2)^2+B_6\theta^2(1-\theta^2)^2+
B_7\theta^2(1-\theta^2)^3\right ]u_{\theta1}=0.
\end{equation}

\begin{figure}[t]
\vspace*{1mm}
\begin{center}
\includegraphics[width=5cm]{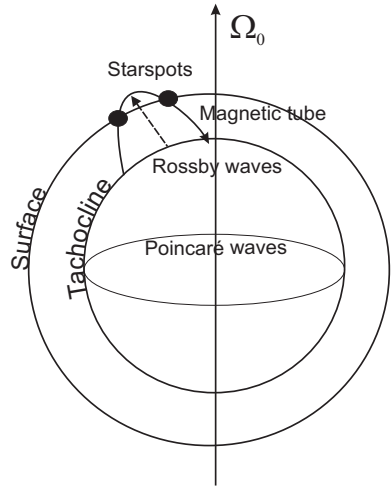}
\end{center}
\caption{Schematic picture of connection between Rossby waves and polar spots in rapidly rotating stars. Magnetic Rossby waves are concentrated near the polar regions in tachoclines of rapidly rotating stars. The erupted magnetic flux tubes lead to the appearance of polar spots at the stellar surfaces.}
\end{figure}

To estimate different terms in this equation, let us consider
$\theta = \epsilon^{-1/4} \eta$, where $\eta$ is close to unity.
Keeping only the terms $\geq \epsilon^{1/2}$ we get from Eq.~\ref{eq:low-weber-mu-theta}
\begin{equation}\label{eq:low-weber-mu-theta1}
{{\partial^2 {u_{\theta1}}}\over {\partial \theta^2}} +
{1\over \theta}{{\partial {u_{\theta1}}}\over {\partial
\theta}} + \left [ - {s^2\over \theta^2} + A_1 - A_2\theta^2\right
]u_{\theta1}=0,
\end{equation}
where
\begin{equation}
A_1= -{{s \lambda}\over {\lambda^2-\alpha^2s^2}}-{{{\epsilon}
\lambda^2 + 3{\epsilon} \alpha^2 s \lambda}\over
{\lambda^2-\alpha^2s^2}} + {{2{\epsilon} \lambda^3}\over
{(\lambda^2-\alpha^2s^2)s}},
\end{equation}
\begin{equation}
A_2= -{{{\epsilon} \lambda^2}\over {\lambda^2-\alpha^2s^2}} +{1\over
4}{{\epsilon^2 \alpha^4 \lambda^2 s^2}\over
{(\lambda^2-\alpha^2s^2)^2}}.
\end{equation}

One can find that Eq.~\ref{eq:low-weber-mu-theta1} is the same as Eq.~\ref{eq:low-weber-theta1} therefore the dispersion relations and
solutions are the same as in the previous section. This means that the magnetic field structure has no significant influence on the dynamics
of magnetic Rossby waves. This is probably because the difference between the two profiles of the magnetic field is most important at the equator but much less important near the poles, where the eigenfunctions of magnetic Rossby waves are concentrated.

\section{Discussion}

Magnetic field and activity determine the structure of stellar chromosphere/corona and have a significant influence on the evolution of exoplanetary atmospheres (Khodachenko et al. \cite{Khodachenko2007}). Therefore, to study the stellar activity/magnetic field is crucial for recently discovered exoplanets. Magnetic activity is connected to starspots, which appear as strong concentrations of magnetic flux at stellar surfaces and thus are similar to sunspots. Emerged magnetic flux is probably connected to deeper regions, namely to the tachocline, which is a thin transition layer between the external convection envelope and the internal radiative core. The tachocline may prevent the spread of the solar angular momentum from the convection zone to the interior (Spiegel and Zahn \cite{Spiegel1992}; Gough and McIntyre \cite{gough98}), and it is probably the place where the large-scale magnetic field that governs the solar/stellar activity is generated/amplified. The magnetic flux rises upwards due to the magnetic buoyancy in the form of tubes and appears as spots at the surface.

Starspots on rapidly rotating stars are observed to be formed everywhere, from low latitudes up to polar regions (Strassmeier \cite{Strassmeier2009}). Enhanced Coriolis force in rapid rotators may push the rising magnetic flux tubes, which begin to rise in lower latitudes as in the solar case, towards higher latitudes to form polar spots (Sch\"ussler and Solanki \cite{Schussler1992}; Sch\"ussler et al. \cite{Schussler1996}; DeLuca et al. \cite{DeLuca1997}; Granzer et al. \cite{Granzer2000}; Holzwarth et al. \cite{Holzwarth2006}, Isik et al. \cite{Isik2007}). But rapidly rotating stars show low latitude spots as well, which is difficult to explain as all tubes should tend to higher latitudes. A meridional flow may also transport the photospheric magnetic flux towards the poles (Schrijver and Title \cite{Schrijver2001}), but the flow should be significantly enhanced in rapid rotators compared to the Sun (Mackay et al. \cite{Mackay2004}).

Besides the longer periodicity caused by the $\alpha-\omega$ or the $\alpha^2$ dynamo ($\sim$ 11 yr for the Sun), stellar magnetic activity also shows shorter periodicity (Massi et al. \cite{massi1998}, \cite{massi2005}, Donati et al. \cite{Donati2008}, Lanza et al. \cite{Lanza2009}). Almost all observed periodicities are much shorter than the solar cycles but longer than the stellar rotation periods. This fact fits well with Rossby-type wave activity. There are observations of intermediate periodicity in solar activity (Rieger et al. \cite{Rieger84}, Carbonell et al. \cite{Carbonell1990}, Gigolashvili et al. \cite{gigolashvili95}, Oliver et al. \cite{oliver1998}, Vecchio et al. \cite{vecchio10}), which were explained by the instability of magnetic Rossby waves (Zaqarashvili et al. \cite{Zaqarashvili2010a}, \cite{Zaqarashvili2010b}). The rapid rotation may enhance the role of large-scale Rossby waves in the dynamics of stellar interior and atmospheres.

Here we study the linear dynamics of shallow water waves in rapidly rotating stellar tachoclines with toroidal magnetic field. The latitudinal differential rotation is not taken into account because it significantly complicates the solutions (Dzhalilov $\&$ Staude \cite{Dzhalilov2004}, Mathis \cite{mathis2009}, Zaqarashvili et al. \cite{Zaqarashvili2010b}). We derive analytical dispersion relations and find the dependence of the wave frequencies on various parameters, which are expressed by angular velocity, reduced gravity, and magnetic field strength. Then we plot the latitudinal structure of the first few harmonics. We find that high-frequency magnetic Poincar\'e waves are located mainly at low latitudes (Section 3.1; Fig. 3). In contrast, low-frequency magnetic Rossby waves are located near the poles in rapid rotators (Section 3.2; Fig. 6). Heng and Spitkovsky (\cite{Heng2009}) also find that the low-frequency waves, called magnetostrophic modes, reside near the poles. Actually, the magnetostrophic modes defined by Heng and Spitkovsky (\cite{Heng2009}) are the same wave modes as the magnetic Rossby waves defined here. The difference between these two considerations is the magnetic field structure. Heng and Spitkovsky (\cite{Heng2009}) used the radial magnetic field, which is valid for neutron star-related applications, while we use the toroidal magnetic field, which is applicable to stellar tachoclines. The latitudinal differential rotation and toroidal magnetic field may lead to the instability of magnetic Rossby waves, which may enhance the magnetic flux in the polar regions. Then the enhanced magnetic flux may rise upwards owing to the magnetic buoyancy and appear as polar spots (see Fig. 7).

The important parameter governing the dynamics of shallow water waves is $\epsilon$ and it should be estimated for rapidly rotating stars. The radius, density, and width of the stellar tachocline can be used from the solar values as $R_{0} = 5\,10^{10}$ \ cm, $\rho = 0.2\, \mathrm {g \, cm^{-3}}$, and $H_0=0.01 \, R_0=5\, 10^{8}\, \mathrm {cm}$, respectively. The subadiabatic temperature gradient of the tachocline may significantly reduce the actual gravity. For example, the reduced gravity at the main, stably stratified part of the solar tachocline can be as low as $5\, 10^2\, \mathrm {cm\, s^{-2}}$ (Zaqarashvili et al. \cite{Zaqarashvili2009}). The typical value of the equatorial angular velocity of the solar tachocline is $\Omega_{\sun}= 2.7\, 10^{-6}\, \mathrm {s^{-1}}$. We assume that $\kappa$ is the ratio between angular velocities of a sample star and the Sun i.e. $\kappa  = \Omega_{0}/\Omega_{\sun}$. Then we have
\begin{equation}\label{eq:epsilon}
\epsilon={{4\Omega^2_{0}R^2_0}\over {g H_0}}\approx 0.3 \, \kappa^2.
\end{equation}
Here $\kappa$ is significantly larger than one for rapidly rotating stars. For example, the young solar analog EK Dra (rotation period of 2.68 days) yields $\kappa \approx 10$, which leads to $\epsilon \approx 30$. More rapidly rotating stars will lead to larger $\epsilon$. Therefore, the approximation of large $\epsilon$ is applicable for most rapid rotators. It should be mentioned that the reduced gravity $g$ and the tachocline thickness $H_0$ are very sensitive parameters (there are no exact values even in the case of Sun) and thus may vary significantly from star to star.

\subsection{Polar spots, young stellar winds, and impact on early planetary environments}

As pointed out in the introduction, young solar-like stars rotate much more faster than the
present Sun. With enhanced high-energy emissions and frequent flares, young stars
are also expected to have more powerful stellar winds (e.g. Wood et al. \cite{Wood2002}, G\"udel 2007).
Wood et al. (\cite{Wood2001}, \cite{Wood2002}) have applied a method to infer stellar wind characteristics from
observations between the interaction of the fully ionized coronal stellar winds and the
partially ionized local interstellar medium. Modeling the
associated Lyman-$\alpha$ absorption features that are formed in the astrospheres of
observed stellar sample stars have provided the first empirical estimate of coronal
mass loss rates for G–K main-sequence stars and indicated that the mass loss rates, so that the expected stellar winds appear to increase with stellar activity.

However, more recent observations by Wood et al. (\cite{Wood2005}) of the
Lyman-$\alpha$ absorption signature of the astrosphere of about
0.55 Gyr old solar-like star $\xi$ Boo indicate that there may
possibly be a high-activity cut-off regarding the mass loss
in a radiation activity relation derived in the form of a power
law by Wood et al. (\cite{Wood2002}). The observations of $\xi$ Boo
indicate that the mass loss of that particular star is about
20 times less than the average stellar mass loss value
inferred from stars with an age of about 3.9 Gyr.

From the findings of our study we would expect a rotation-related modification
of the outflow region for young stellar winds, which may have affected the
observation of Wood et al. (\cite{Wood2002}) in the equatorial plane. Polar starspots may completely
change the structure of stellar winds in the rapidly rotating, young solar analogs.
It is known that the dense slow solar wind is initiated above active regions with
closed magnetic field lines at low latitudes, while the fast solar wind originates
mostly in high-latitude coronal holes with opened field lines. The polar appearance of
starspots may force the slow stellar wind to flow from the polar regions, so almost
parallel to the rotation axis. Then this may change the current views about the magnetic
braking of stellar rotation in the young age of their evolution.

Stellar magnetic activity has a tremendous influence on the planetary evolution. Coronal mass
ejections in young, more active stars may lead to the erosion of atmospheres in exoplanets
(Khodachenko et al. \cite{Khodachenko2007}; Lammer et al. \cite{Lammer2007}). On the other hand,
the polar spots in young stars could change both the structure of coronal magnetic fields and the formation of prominences, which are the actual cause of CMEs. The polar location of
prominences may trigger polar CMEs, which propagate significantly out of the ecliptic plane.
Then the propagation of shocks and magnetic ropes/clouds out of the ecliptic plane may completely
change the current picture of stellar-planetary interaction. The polar spots and related enhanced plasma flow out of the ecliptic plane during the fast-rotation
period of the young Sun, or similar stars may lead to a
longer stability of atmospheres related to planets that orbit in the ecliptic plane, and
therefore to more convenient conditions of planetary evolution. Although our findings
agree with the mass-loss observations of Wood et al. (\cite{Wood2005}), clearly more measurements are needed to better define the mass-loss,
radiation activity relation of cool main-sequence stars, at high activity levels.
More observations of G and K-type stars with X-ray
surface fluxes greater than 10$^6$ erg cm$^{-2}$ s$^{-1}$ are needed to
find out what is happening to stellar plasma outflow during
high coronal activity periods. All these problems need to be studied in more detail
in the future.

\section{Conclusions}

Large-scale magnetic Rossby waves may play a significant role in the dynamics of rapidly rotating stars. Here we have studied the shallow-water MHD waves in the tachoclines of rapid rotators. We considered a rotating spherical coordinate system and a toroidal magnetic field in the tachocline. We derived the dispersion relations for high (magnetic Poincar\'e waves) and low (magnetic Rossby waves) frequency spectra separately in the approximation of rapid rotation. We found that magnetic Rossby waves tend to locate mainly at the polar regions, while magnetic Poincar\'e waves
are located at low latitudes. Magnetic Rossby waves may lead to the enhancement of magnetic flux at the polar regions and trigger its eruption towards the surface in the form of magnetic flux tubes. Then the magnetic tubes may form starspots at the polar region. Magnetic Rossby waves may cause the observed intermediate period oscillations in the magnetic activity of rapidly rotating stars, such as young solar proxies. The polar location of
prominences during the fast-rotating period of the young star may trigger polar CMEs, which propagate
significantly out of the ecliptic plane and completely change the current picture of the solar/stellar plasma
interaction with young planetary environments.

\begin{acknowledgements}
The work was supported by the Austrian Fonds zur F\"orderung der
wissenschaftlichen Forschung (projects P21197-N16 and P22950-N16). T.V.Z. acknowledges financial support from the Georgian National Science
Foundation (under grant GNSF/ST09/4-310). R.O., M.C. and J.L.B. acknowledge financial support from grant AYA2006-07637 and Feder funds. The authors thank the referee, Dr. S. Mathis, for stimulating comments.

\end{acknowledgements}


\begin{thebibliography}{}

\bibitem[1964]{abramowitz}Abramowitz, M., \& Stegun, I.A. 1964,
{\it Handbook of Mathematical Functions} (Washington, D.C.: National
Bureau of Standards)
\bibitem[2005]{Arfken2005}Arfken, G.B. \& Weber, H.J., 2005, {\it Mathematical Methods for Physicists} (Burlington: Elsevier Acadenic Press)
\bibitem[1997]{Ayres1997}Ayres, T. R., 1997, \jgr, 102, 1641
\bibitem[1998]{Barnes1998}Barnes, J. R., Collier Cameron, A., Unruh, Y.C., Donati, J.F. and Hussain, G.A.J., 1998, \mnras, 299, 904
\bibitem[2005]{Berdyugina2005}Berdyugina, S.V., 2005, Living Rev. Solar Phys. 2, 8
\bibitem[1990]{Carbonell1990} Carbonell, M., \& Ballester,
J. L. 1990, \aap, 238, 377
\bibitem[1997]{DeLuca1997}DeLuca, E., Fan, Y., \& Saar S., 1997, \apj, 481, 369
\bibitem[1992]{Donati1992}Donati, J.F., Brown, S.F., Semel, M., Rees, D.E., Dempsey, R.C., Matthews, J.M., Henry, G.W., and Hall, D.S., 1992, \aap, 265, 682
\bibitem[1999]{Donati1999}Donati, J.F., Collier Cameron, A., Hussain, G.A.J. \& Semel, M., 1999, \mnras, 302, 437
\bibitem[2003]{Donati2003}Donati, J.F., Collier Cameron, A., Semel, M., et al., 2003, \mnras, 345, 1145
\bibitem[2008]{Donati2008}Donati, J.F., Moutou, C., Fares, R., et al., 2008, \mnras, 385, 1179
\bibitem[1972]{Durney1972}Durney, B. 1972, in Solar Wind, ed. C. P. Sonett, P. J. Coleman, \& J. M. Wilcox (Washington: NASA), 282
\bibitem[2004]{Dzhalilov2004}Dzhalilov, N.S. $\&$ Staude, J., 2004, \aap, 421, 305
\bibitem[1995]{gigolashvili95}Gigolashvili, M. Sh., Japaridze, D. R., Pataraya, A. D. \& Zaqarashvili, T.V., 1995, \solphys, 156, 221
\bibitem[1982]{gill82}Gill, A. E. 1982, Atmosphere-Ocean Dynamics (San Diego: Academic Press)
\bibitem[1997]{gilman97} Gilman, P. A. $\&$ Fox, P. A. 1997, \apj, 484, 439
\bibitem[2000]{Gilman2000}Gilman, P.A. 2000, \apj, 484, 439
\bibitem[1998]{gough98} Gough, D.O. $\&$ McIntyre, M.E. 1998, \nat, 394, 755
\bibitem[2000]{Granzer2000}Granzer, Th., Sch\"ussler, M., Caligari,  P. and Strassmeier,
K.G. 2000, \aap, 355, 1087
\bibitem[1997]{Gudel1997}G\"udel, M., Guinan, E.F. $\&$ Skinner, S.L., 1997, \apj, 483, 947
\bibitem[2007]{Khodachenko2007}Khodachenko, M.L. et al., 2007, Astrobiology, 7,
167
\bibitem[2006]{Kulikov2006}Kulikov Y. N., Lammer, H., Lichtenegger, H.I.M.,  et al., 2006, Planet. Space Sci., 54, 1425
\bibitem[1972]{Hall1972}Hall, D.S., 1972, \pasp, 84, 498, 323
\bibitem[2009]{Heng2009}Heng, K. and Spitkovsky, A., 2009, \apj, 703, 1819
\bibitem[2006]{Holzwarth2006}Holzwarth, V., Mackay, D.H., and Jardine, M., 2006, \mnras, 369, 1703
\bibitem[2007]{Isik2007}Isik, E., Sch\"ussler, M. and Solanki, S.K.,
2007, \aap, 464, 1049
\bibitem[2007]{Lammer2007}Lammer, H., Lichtenegger, H.I.M., Kulikov, Y.N., et al., 2007, Astrobiol., 7, 185
\bibitem[2011]{Lammer2011}Lammer, H.,  G\"udel, M., Kulikov, Y. N.,  et al., 2011, Earth Planet. Space (accepted)
\bibitem[2009]{Lanza2009}Lanza, A.F., Pagano, L., Leto, G., et al., 2009, \aap, 493, 193
\bibitem[1965]{Longuet-Higgins1965}Longuet-Higgins, M.S., 1965, Proceedings of The Royal Society Series / A, 284, 40
\bibitem[1968]{Longuet-Higgins1968} Longuet-Higgins, M. S. 1968, Proc. R. Soc. London. A, 262, 511
\bibitem[2004]{Mackay2004}Mackay, D.H., Jardine, M., Collier Cameron, A.,
Donati, J.F., Hussain, G.A.J., 2004, \mnras, 354, 737
\bibitem[1998]{massi1998} Massi, M., Neidh\"{o}fer, J., Torricelli-Ciamponi, G. $\&$
Chiuderi-Drago, F. 1998, \aap, 332, 149
\bibitem[2005]{massi2005} Massi, M., Neidh\"{o}fer, J., Carpentier, Y. $\&$ Ros, E. 2005, \aap, 435,
L1
\bibitem[2009]{mathis2009}Mathis, S., 2011, \aap, 506,
811
\bibitem[2011]{mathis2011}Mathis, S. $\&$ de Brye, N. 2011, \aap, 526,
A65
\bibitem[2009]{Olah2009}Ol\'ah, K., Koll\'ath, Z., Granzer, T.,  et al., 2009, \aap, 501, 703
\bibitem[1998]{oliver1998} Oliver, R., Ballester, J. L.,
\& Baudin, F. 1998, \nat, 394, 552
\bibitem[2005]{Ribas2005}Ribas, I., Guinan, E. F., G\"udel, M. \& Audard, M., 2005, \apj, 622, 680
\bibitem[1984]{Rieger84}Rieger, E., Share, G. H., Forrest,
D. J., Kanbach, G., Reppin, C., \& Chupp, E. L. 1984, Nature, 312,
623
\bibitem[1981]{sakurai81}Sakurai, K., 1981, \solphys, 74, 35
\bibitem[2001]{Schrijver2001}Schrijver, K.J. \& Title, A.M., 2001,
\apj, 551, 1099
\bibitem[1992]{Schussler1992}Sch\"ussler, M. and Solanki, S.A., 1992, A\&A, 264, 13
\bibitem[1996]{Schussler1996}Sch\"ussler, M., Caligari, P., Ferriz-Mas, A., Solanki, S.A. and Stix,
M., 1996, A\&A, 314, 503
\bibitem[1985]{Simon1985}Simon, T., Boesgaard, A.M., Herbig, G., 1985, \apj, 293, 551
\bibitem[1972]{Skumanich1972}Skumanich, A., 1972, \apj, 171, 565
\bibitem[1982]{Soderblom1982}Soderblom, D. R., 1982, \apj, 263, 239
\bibitem[1992]{Spiegel1992}Spiegel, E.A. \& Zahn, J.-P.
1992,\aap, 265
\bibitem[1990]{Strassmeier1990}Strassmeier, K.G., 1990, \apj, 348, 682
\bibitem[2009]{Strassmeier2009}Strassmeier, K.G., 2009, \aap Rev., 17, 251
\bibitem[2010]{vecchio10}Vecchio, A., Laurenza, M., Carbone, V. \& Storini, M., 2010, \apj, 709, L1
\bibitem[1983]{Vogt1983}Vogt, S.S. and Penrod, G.D., 1983, \pasp, 95, 565
\bibitem[2001]{Wood2001}Wood, B. E., Karovska, M., Hack, W., 2001, \apj, 556, L51
\bibitem[2002]{Wood2002}Wood, B.E., M\"uller, H.-R., Zank, G. P., Linsky, J. L., 2002, \apj, 574, 412
\bibitem[2005]{Wood2005}Wood, B.E., M\"uller, H.-R., Zank, G. P., Linsky, J. L., Redfield, S., 2005, \apj, 628, L143
\bibitem[2007]{Zaqarashvili2007}Zaqarashvili, T. V., Oliver, R., Ballester, J. L. \& Shergelashvili, B. M. 2007, \aap, 470, 815
\bibitem[2009]{Zaqarashvili2009}Zaqarashvili, T. V., Oliver, R. \&  Ballester, J. L. 2009, \apj, 691, L41
\bibitem[2010a]{Zaqarashvili2010a}Zaqarashvili, T. V., Carbonell, M., Oliver, R. \& Ballester, J. L. 2010a, \apj, 709, 749
\bibitem[2010b]{Zaqarashvili2010b}Zaqarashvili, T. V., Carbonell, M., Oliver, R. \& Ballester, J. L. 2010b, \apjl, 724, L95

\end{thebibliography}
\end{document}